\documentclass[nofootinbib,preprint,amsmath,showkeys,showpacs,amssymb,aps,longbibliography]{revtex4-1}
\usepackage{graphicx}
\usepackage[utf8,latin1]{inputenc}
\usepackage{dcolumn}
\usepackage{forloop}
\usepackage{xcolor}
\usepackage{xpatch}
\usepackage{blindtext}
\usepackage{bm}
\usepackage{footmisc}
\usepackage{footnotehyper}
\usepackage{leftidx}
\usepackage{hyperref}
\usepackage{framed}
\usepackage{mathtools}
\usepackage{amsmath}

\usepackage{tcolorbox}
\usepackage{tikz-cd}
\newcommand{\qm}[1]{``#1''}
\usepackage{hyperref}
\hypersetup{colorlinks, linkcolor={red},citecolor={blue},urlcolor={blue}}  
\newcounter{loopcntr}

\newcommand{\bsubeqs}{\begin{subequations}}
\newcommand{\esubeqs}{\end{subequations}}

\begin{document}

\title[Nonsingular bouncing cosmology in general relativity: physical analysis of the spacetime defect]{Nonsingular bouncing cosmology in general relativity: \\ physical analysis of the spacetime defect}

\author{Emmanuele Battista$^{1,2}$\vspace{0.5cm}}\email{emmanuele.battista@kit.edu; emmanuelebattista@gmail.com}

\affiliation{$^1$ Institute for Theoretical Physics, Karlsruhe Institute
of Technology (KIT), 76128 Karlsruhe, Germany\\
$^2$ Institute for Astroparticle Physics, Karlsruhe Institute of Technology (KIT), Hermann-von-Helmholtz-Platz 1, 76344 Eggenstein-Leopoldshafen, Germany }

\date{\today}

\nopagebreak

\begin{abstract}

In this paper, we describe physical effects occurring in the regularized Robertson-Walker spacetime which can reveal the presence of the defect.  Our analysis is based on two main physical quantities: the compressive forces acting  on (human) observers and the energy possessed by massive particles and photons during their dynamical evolution. In Sec. \ref{Sec:Compressive_forces}, we claim that with a characteristic length scale of the order of the Planck length compressive forces become so intense near the defect that no (human) observer is able to cross it. In Sec. \ref{Sec:Energy}, we show that the energy exhibits an unusual character over a small time interval around the bounce contrasting with the behaviour in the  standard cosmology picture. We conclude the paper with some considerations and open problems related to our results.
\end{abstract}

\nopagebreak
\maketitle

\section{Introduction}\label{Sec:Introduction}

Spacetime defect has been recently proposed in the literature as a tool  to tame the big-bang singularity \cite{KI,KII,KWI,KWIII}. Such an object can be described by a degenerate metric with a vanishing determinant on a 3-dimensional submanifold of the spacetime and a nonzero length scale $b$, which acts as a \qm{regulator} of the Friedmann singularity. This new metric has been called the regularized-big-bang metric.  It gives rise to a nonsingular spatially flat Friedmann-type solution of the Einstein gravitational field equation which allows for a \qm{pre-big-bang} phase with a bounce-type behaviour of the cosmic scale factor. Cosmological observables occurring in this geometry such as past particle horizon and modified Hubble diagrams have been investigated in Ref. \cite{KWI} and the effective violation of the null energy condition (NEC)  in the vicinity of the defect was first pointed out in Ref. \cite{KII} (for details see Ref. \cite{KWIII}).   A gravitational model for the nonsingular bounce solution involving a Brans-Dicke-type scalar field having, in the potential action, a \qm{wrong-sign} kinetic component and a quartic interaction term has been presented in Ref. \cite{Klinkhamer:2020sxv}, where it is shown that the bounce behaviour appears if the boundary conditions provide for a kink-type solution of the Brans-Dicke-type scalar field.

The  physical investigation of  the defect is a  subtle issue, since classical physics may not be valid at  $t=0$.  Possible connections between the  characteristic length scale $b$  and the Planck length suggested by loop quantum cosmology \cite{Ashtekar2008} and string cosmology \cite{Lidsey1999} have been explored in  Appendix B of Ref. \cite{KII} (see also Appendix C of Ref. \cite{KWIII}). Furthermore, a new explanation about the origin of $b$ has been set forth recently in Ref. \cite{K2020-IIB-3} (see also Refs. \cite{Klinkhamer:2020wct,Klinkhamer2021a}),  where  it has been shown that the classical regularized-big-bang metric can, in principle, emerge from  the IIB matrix model (i.e., a nonperturbative formulation of type-IIB superstring theory \cite{Becker2006}).  This means that physics of the spacetime defect  might require  knowledge of the underlying model and quantities at $t=0$ might originate from the underlying theory. The situation might be  similar to that of an atomic crystal, where classical physics is a reliable source of information everywhere except at  the atomic defect, whose details necessitate quantum mechanics. In the same way, according to the picture of Ref. \cite{K2020-IIB-3}, the classical spacetime can be emergent for any value of the $t$ variable and the \qm{point} $t=0$ is only defined via a limit procedure.

Motivated by the fact that the description of physics at $t=0$ is a delicate point, in this paper we will examine physical effects pointing out the presence of the defect in the regularized Robertson-Walker (RW) spacetime. We will see that if the length scale $b$ had a quantum nature, then the defect would be shaped as an object allowing no human observer to go across it. Furthermore, we will explain how massive particles and photons energy display, in the proximity of the defect, a behaviour deviating from the expectations of standard RW cosmology. 

The plan of the paper is as follows. In Sec. \ref{Sec:Freely_falling_obs_FLRW}, we recall some basic concepts of modified RW spacetime and define two different freely falling observers which will fulfil a crucial role in our analysis:  the Eulerian observer and the non-comoving observer, which we will call the \qm{traveller}.  In Sec. \ref{Sec:Compressive_forces}, we will investigate how compressive forces affect these observers. For this purpose, we will suppose to deal with human observers, i.e., observers made of atoms. In Sec. \ref{Sec:Energy}, we propose a definition of energy suitable for our model and analyze its features. Eventually, concluding remarks are made in Sec. \ref{Sec:Conclusions}.

Throughout the paper, we use  metric signature $(-,+,+,+)$ and natural units with $c=1$ and $\hbar=1$.

\section{Freely falling observers in the regularized Robertson-Walker geometry} \label{Sec:Freely_falling_obs_FLRW}

In Refs.  \cite{KI,KII} it has been shown that the big-bang singularity underlying the standard Friedmann cosmology can be regularized by employing the following \emph{Ansatz} for the modified spatially flat RW metric:
\bsubeqs \label{modified_FRW}
\begin{equation} \label{modified_line_element}
{\rm d}s^2 = -\dfrac{t^2}{t^2+b^2} {\rm d}t^2 + a(t)^2 \delta_{ij} {\rm d}x^i {\rm d}x^j, \quad \quad (i,j=1,2,3),
\end{equation}
\begin{equation} 
b^2>0,
\end{equation}
\begin{equation}
a(t) \in \mathbb{R},
\end{equation}
\begin{equation} \label{t_coord}
t \in \left(-\infty, +\infty\right),
\end{equation}
\begin{equation}\label{x^i_coord}
x^i \in \left(-\infty, +\infty\right),
\end{equation}
\esubeqs
where $t$ denotes the cosmic time coordinate, $\{x^1,x^2,x^3\}$ the comoving spatial Cartesian coordinates, $a(t)$ the cosmic scale factor and $b>0$ corresponds to the characteristic length scale of the spacetime defect localized at $t=0$.  By employing the metric (\ref{modified_FRW}) and the energy-momentum tensor of a homogeneous perfect fluid having energy density $\rho$ and pressure $P$, the Einstein equation with a vanishing cosmological constant gives \cite{KI,KII}
\bsubeqs\label{modified-Friedmann-eqs}
\begin{equation} \label{Friedmann-1-order}
\left[1+\dfrac{b^2}{t^2} \right] \left(\dfrac{\dot{a}}{a}\right)^2 = \dfrac{8 \pi G}{3} \rho,
\end{equation}
\begin{equation}\label{Friedmann-2-order}
\left[1+\dfrac{b^2}{t^2} \right]  \left[\dfrac{\ddot{a}}{a}+\dfrac{1}{2}\left(\dfrac{\dot{a}}{a} \right)^2\right]-\dfrac{b^2}{t^3} \dfrac{\dot{a}}{a} = -4 \pi G P,
\end{equation}
\begin{equation} \label{energy-conservation-eq}
\dot{\rho} +3 \dfrac{\dot{a}}{a} \left(\rho + P \right)=0,
\end{equation}
\begin{equation}\label{equation-of-state}
P=P\left(\rho\right),
\end{equation}
\esubeqs
where  the dot signifies a differentiation with respect to the $t$ variable. Equations (\ref{Friedmann-1-order}) and (\ref{Friedmann-2-order}) are the modified   first-order and second-order spatially flat Friedmann equations, respectively,  Eq. (\ref{energy-conservation-eq}) the energy-conservation equation of the matter, and Eq. (\ref{equation-of-state}) the equation of state. Two remarks are in order. First, since the inverse metric from (\ref{modified_line_element}) has the  $g^{00}$ component which diverges at $t=0$, the reduced field equations at $t=0$ must be obtained carefully from the limit $t \to 0$ (see Sec. 3.3.1 of Ref. \cite{Gunther} for further details). Second, the new $b^2$ terms in the modified Friedmann equations  (\ref{Friedmann-1-order}) and (\ref{Friedmann-2-order}) are a manifestation of the different differential structure of (\ref{modified_line_element}) compared to the differential structure of the standard spatially flat RW metric \cite{Wald,MTW} which gives the standard Friedmann equations (see Refs. \cite{KI,KII,KWI} for details). 

From Eq. (\ref{modified-Friedmann-eqs}), we obtain for the function $a(t)$ the following solutions (with normalization $a(t_0)=1$ at $t_0>0$) \cite{KI,KII}:
\begin{equation} \label{eq:scalefactor}
a(t) = \left \{ 
\setlength{\tabcolsep}{10pt} 
\renewcommand{\arraystretch}{1.8}
\begin{array}{rl}
& \left(\dfrac{b^2+t^2}{b^2 + t_0^2}\right)^{1/3}, \quad {\rm nonrelativistic\; matter},\\
& \left(\dfrac{b^2+t^2}{b^2 + t_0^2}\right)^{1/4}, \quad {\rm relativistic\; matter},
\end{array}
\right.
\end{equation}
depending on whether a   relativistic-matter or a nonrelativistic-matter equation of state for the perfect fluid is used. From Eq. (\ref{eq:scalefactor}), the  bouncing behavior of the positive scale factor  is evident: $a(t)$ decreases (resp. increases) for negative (resp. positive) $t$ and the bounce occurs at $t = 0$ where $a(t)$ has a vanishing time derivative. 

For later use, we also give the affine connection component
\begin{equation} \label{eq:Gamma_sec_2}
\Gamma^t_{\phantom{t}tt}=\dfrac{b^2}{t\left(b^2 + t^2\right)}.
\end{equation}

One of the main advantages of our model consists in the fact  that it depends on only one free parameter, i.e., the defect length scale $b$. In this paper,  we will examine the effects produced by  the possible range of values  of $b$. In that regard, we recall that in Ref. \cite{KWI} a \emph{Gedankenexperiment} which makes use of  modified Hubble diagrams has been proposed in order to evaluate   $b$ numerically.  Furthermore, it should be stressed that the study of cosmological perturbations is a precious mean to determine observable constraints on $b$. Indeed, this analysis makes it possible to determine several observables  like the spectral index of the primordial curvature perturbations, the  tensor-to-scalar ratio, and the running of the spectral index, permitting thus  a comparison with  the latest Planck 2018 data \cite{Planck2018} (for a thorough investigation of cosmological perturbations framed  in extended theories of gravity we refer the reader to \cite{Odintsov2019,Odintsov2020a,Odintsov2020b} and references therein). Scalar metric perturbations for the modified RW geometry (\ref{modified_FRW}) have been tackled  in Ref. \cite{KWIII}\footnote{Vector and tensor perturbations are also briefly analyzed in Appendix B of Ref. \cite{KWIII}.},  where it has been proved the stability of the bounce  under small perturbations of the metric and the matter. Likewise modified Friedmann equations (\ref{modified-Friedmann-eqs}), which represent singular differential equations having  nonsingular solutions \cite{KI,KII}, the metric perturbations exhibit nonsingular solutions, although they are described by singular differential equations (the singularity appears at $t=0$) \cite{KWIII}. However, it is worth mentioning  that   since the pre-bounce contracting phase is unstable to the growth of anisotropies,  our model is plagued by the Belinskii- Khalatnikov-Lifshitz (BKL) instability \cite{BKL1970}.

In the study of  the generation era of the primordial  perturbation modes  from Bunch-Davies vacuum state  \cite{Bunch1978,Birrell1982}, a   key parameter is represented by the comoving Hubble radius, which is defined as 
\begin{equation} 
R_{\rm H} (t)= \dfrac{1}{a(t) H(t)},
\end{equation}
 $H(t) = \dot{a}(t)/a(t)$ being the Hubble rate. The scale factor  (\ref{eq:scalefactor}) leads immediately  to the following expressions:
\begin{equation} \label{eq:comoving-Hubble-radius}
R_{\rm H} (t) = 
\left \{ 
\setlength{\tabcolsep}{10pt} 
\renewcommand{\arraystretch}{1.8}
\begin{array}{rl}
& \dfrac{3\left(b^2+t^2\right)^{2/3}\left(b^2+t_0^2\right)^{1/3}}{2t}, \quad {\rm nonrelativistic\; matter},\\
& \dfrac{2\left(b^2+t^2\right)^{3/4}\left(b^2+t_0^2\right)^{1/4}}{t}, \quad {\rm relativistic\; matter}.
\end{array}
\right.
\end{equation}
\begin{figure*}[th!]
\centering
\includegraphics[scale=0.74]{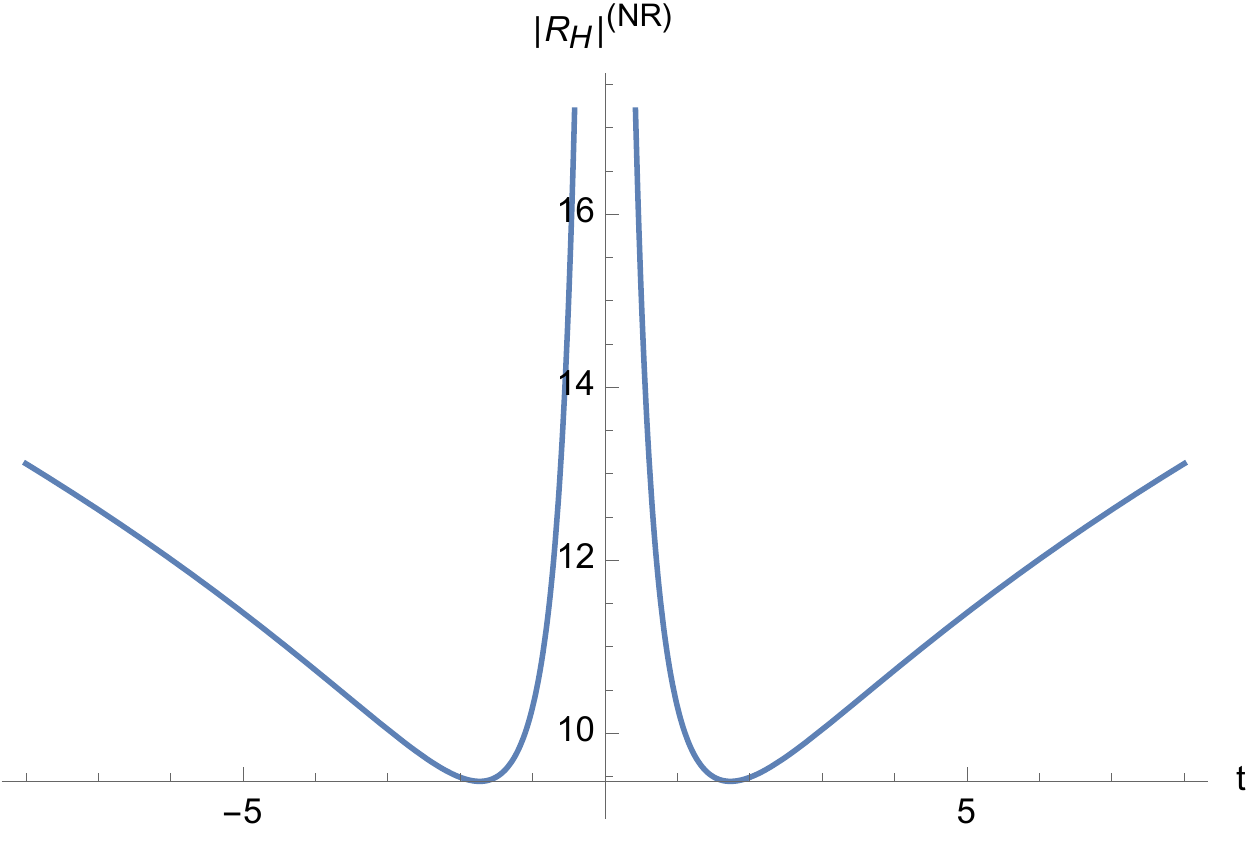}
\caption{The absolute value of the  comoving Hubble radius (\ref{eq:comoving-Hubble-radius}) for the nonrelativistic-matter solution  and with $b=1$ and $t_0=4 \sqrt{5}$. The suffix \qm{NR} stands for \qm{nonrelativistic}  cosmological matter content.}
\label{fig:Comoving_Hubble_radius_non_relat}
\end{figure*}
\begin{figure*}[th!]
\centering
\includegraphics[scale=0.74]{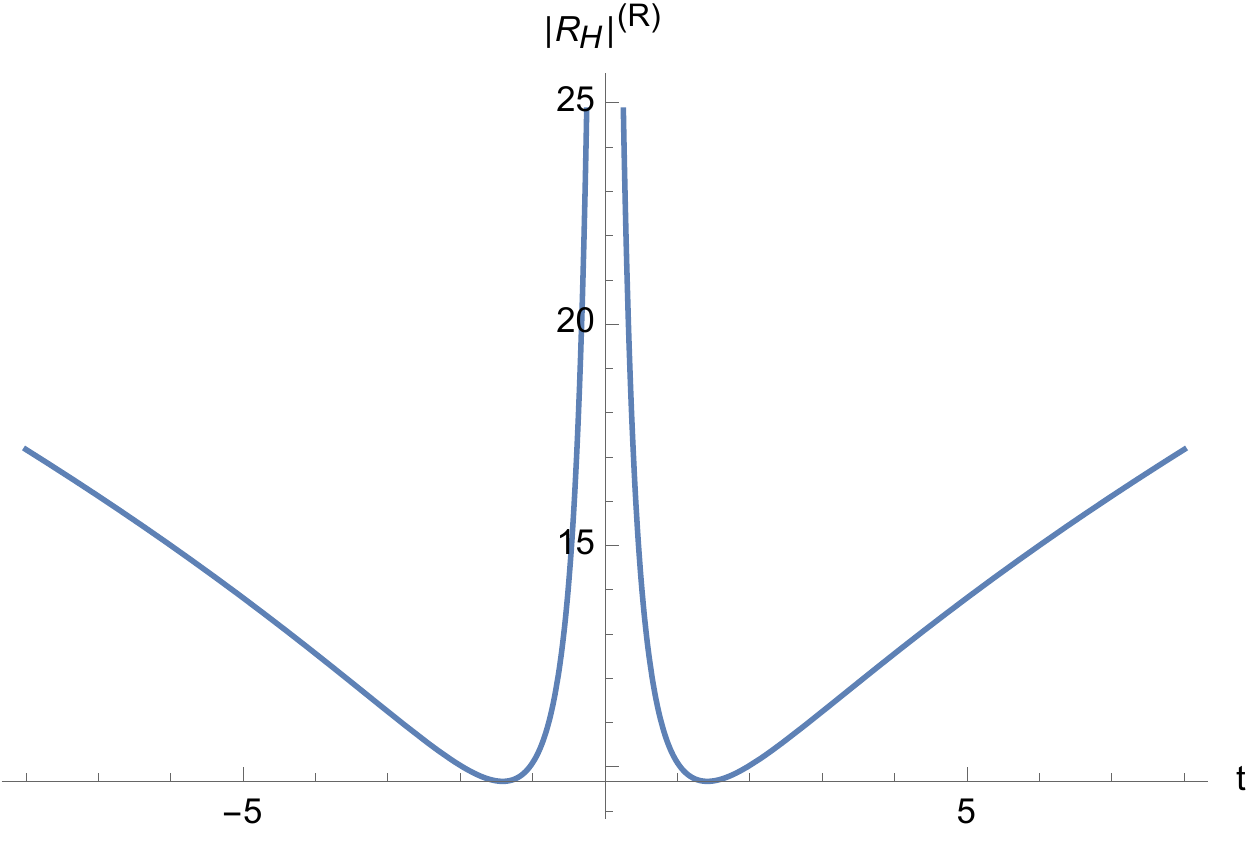}
\caption{The absolute value of the comoving Hubble radius (\ref{eq:comoving-Hubble-radius}) for the relativistic-matter solution  and with $b=1$ and $t_0=4 \sqrt{5}$. The suffix \qm{R} stands for \qm{relativistic}  cosmological matter content.}
\label{fig:Comoving_Hubble_radius_relat}
\end{figure*}
The plot of (the absolute value of) $R_{\rm H}(t) $ is displayed in Figs. \ref{fig:Comoving_Hubble_radius_non_relat} and \ref{fig:Comoving_Hubble_radius_relat}. It is clear that  the comoving Hubble radius diverges both at $t=0$
(which is a peculiar feature of all bouncing cosmologies) and, most importantly, at great distances from the defect. 
Therefore,  the asymptotic behaviour of  the comoving Hubble radius (\ref{eq:comoving-Hubble-radius}) is the same as the one  predicted by those bouncing models  characterized by  primordial perturbation modes  generated at very
large negative cosmic times (see e.g. \cite{Odintsov2019}). The other viable picture occurs when $R_{\rm H}(t) $ vanishes   asymptotically, resulting in  perturbation modes produced near the bouncing epoch (see Ref. \cite{Odintsov2020b} for further details).

In view of our forthcoming analysis, it will be crucial    to define  two different types of observers. First of all, we have  the freely falling \emph{comoving} observer known as  the Eulerian observer. It is easy to show that for this  observer, whose proper time will be indicated with $\tau$, the unit timelike four-velocity vector can be written as \cite{MTW}
\begin{equation} \label{eq:Eulerian_velocity}
\dfrac{{\rm d}x^\alpha}{{\rm d}\tau} \equiv n^\alpha = \dfrac{\sqrt{t^2+b^2}}{\vert t \vert}(1,0,0,0),
\end{equation}
where we have introduced the usual notation $x^\mu=\left(t,x^i\right)$. It will soon be clear that  the motion of the Eulerian observer is characterized by a vanishing conserved momentum $\Pi$ (see Eq. (\ref{eq:def_of_Pi}) below). We also note that Eq. (\ref{eq:Eulerian_velocity}) is not defined at $t=0$ (cf. Eq. (\ref{t_coord})). However, we will see that this fact will not prevent us from computing quantities we are interested in.

The second observer, which throughout the paper will be referred  to as the \qm{traveller}, is a generic  freely falling \emph{non-comoving} observer  having a unit timelike four-velocity vector given by
\begin{equation}\label{Eq:def_traveller_vel}
\dfrac{{\rm d}x^\alpha}{{\rm d}\tau^\prime} \equiv v^\alpha,
\end{equation} 
where $\tau^\prime$ denotes the traveller's proper time. Due to the spatial maximal symmetry of (\ref{modified_line_element}) and the associated conserved angular momentum, we will  suppose that, without loss of generality,  the traveller moves in the $(t,x)$-plane. Therefore, from the condition $\boldsymbol{v} \cdot \boldsymbol{v} =-1$ we obtain
\begin{equation}
\left(\dfrac{{\rm d}t}{{\rm d}\tau^{\prime}}\right)^2 = \left(\dfrac{t^2+b^2}{t^2}\right) \left[1+a(t)^2 \left( \dfrac{{\rm d}x}{{\rm d}\tau^{\prime}}\right)^2 \right].
\end{equation}
Furthermore, we can define the conserved momentum $\Pi$ along the traveller's timelike geodesic as
\begin{equation} \label{eq:def_of_Pi}
\Pi =  V^{(x)}_\alpha  \dfrac{{\rm d}x^\alpha}{{\rm d}\tau^\prime} = a(t)^2 \dfrac{{\rm d}x}{{\rm d}\tau^\prime},
\end{equation}
$\boldsymbol{V}^{(x)}$ being the spacelike $x$-translational Killing vector. 
Therefore, the nonvanishing components of the traveller's four velocity (\ref{Eq:def_traveller_vel}) read as
\bsubeqs \label{eq:traveler_geodesic}
\begin{equation} \label{eq: dt/dtau_prime}
v^t   = \sqrt{\left(\dfrac{t^2+b^2}{t^2}\right) \left(1+\dfrac{\Pi^2}{a(t)^2} \right)},
\end{equation}
\begin{equation}
v^x =  \dfrac{\Pi}{a(t)^2}. 
\end{equation}
\esubeqs
Bearing in mind the above equations, if we consider a setup where the traveller moves along the increasing direction of the $x$-axis (i.e., $\Pi>0$ and ${\rm d}x/{\rm d}t>0$), then we  can write the corresponding geodesic equation as
\begin{equation} \label{eq:traveller_timelike_geod_eq}
\dfrac{{\rm d} x(t)}{{\rm d}t}= \dfrac{\vert t \vert}{\sqrt{t^2 + b^2}} \dfrac{1}{\sqrt{a(t)^2 \left(1 + a(t)^2/\Pi^2\right)}}.
\end{equation}
We have solved Eq. (\ref{eq:traveller_timelike_geod_eq}) and  have seen that the traveller's  timelike geodesics turn out to be well-behaved at $t=0$. This agrees with the analysis of \emph{null} geodesics performed in Sec. IIIA of Ref. \cite{KWI}. However, it should be stressed that,   since Levi-Civita connection of degenerate metrics is not unique,  some  ambiguities in the study of geodesic equation can arise (see Ref. \cite{Gunther} for further details).

\section{Compressive forces acting on  observers} \label{Sec:Compressive_forces}

In this section, we will evaluate the compressive forces felt by the Eulerian observer and the traveller as they travel towards the defect. In that regard, we will suppose to deal with human observers, meaning that we will assume that they are made of atoms.

In order to evaluate the abovementioned compressive forces, we will employ two different instantaneous rest frames:  the proper reference frame $\bold{e}_{\hat{\alpha}} = (\bold{e}_{\hat{\tau}},\bold{e}_{\hat{x}},\bold{e}_{\hat{y}},\bold{e}_{\hat{z}})$ of the Eulerian observer and the proper reference frame $\bold{e}_{\hat{\alpha}^{\prime}}=(\bold{e}_{\hat{\tau}^{\prime}},\bold{e}_{\hat{x}^{\prime}},\bold{e}_{\hat{y}^{\prime}},\bold{e}_{\hat{z}^{\prime}})$ of the traveller. Such frames will be constructed in Eqs. (\ref{eq:cappuccio_first})--(\ref{eq: e_alpha_cappuccio}) and (\ref{eq:boost_velocity}) and (\ref{eq:e-alpha_cappuccio_prime}), respectively.
We are aware of the fact  that in our model  (\ref{modified_FRW}) the equivalence principle does not hold at $t=0$ and for this reason we will employ the limit $t \to 0$ to evaluate at $t=0$ the quantities we are interested in  (details can be found in Ref. \cite{Gunther}).

The compressive forces felt by a (human) observer are measured via the components of the Riemann curvature tensor occurring in the geodesic deviation equation evaluated in his/her local orthonormal frame. We recall that the geodesic deviation equation can be written as \cite{MTW}
\bsubeqs\label{eq:geodesic_deviation_equation}
\begin{equation}
\Delta a^\alpha = - F^{\alpha}_{\rm compressive},
\end{equation}
\begin{equation}
F^{\alpha}_{\rm compressive} = R^{\alpha}_{\phantom{\alpha}\beta\gamma\delta} u^\beta \xi^\gamma u^\delta,
\end{equation}
\esubeqs
where $\Delta \boldsymbol{a} = \nabla_{\boldsymbol{u}} \nabla_{\boldsymbol{u}}\boldsymbol{\xi} $ is the relative acceleration of two freely falling test particles (i.e., two nearby geodesics) having separation vector  $\boldsymbol{\xi}$ and  four-velocity $\boldsymbol{u}$.

With the above premises, we are ready to construct the proper reference frame $\{\bold{e}_{\hat{\alpha}}\}$ of the Eulerian observer  and to calculate compressive forces acting on him/her. We will see that we can get the traveller's proper reference frame $\{\bold{e}_{\hat{\alpha}^{\prime}}\}$ by simply applying  a Lorentz boost to Eulerian observer's frame. Since   both the Eulerian observers and the traveller are freely falling observers, their proper reference frames turn out to be local Lorentz frames all along their geodesics worldline (with the exclusion of  $t=0$, i.e., the defect). The employed coordinates (which are Riemann normal coordinates with axes marked by gyroscopes, see Sec. 13.6 of Ref. \cite{MTW} and also Refs.  \cite{Hartle2003gravity,Manasse-Minser1963} for details) are know as Fermi normal coordinates (we have explicitly checked that the conditions $\nabla_{\boldsymbol{n}}\bold{e}_{\hat{\alpha}}=0$ and $\nabla_{\boldsymbol{v}}\bold{e}_{\hat{\alpha}^{\prime}}=0$ hold). By invoking the usual tetrad formalism \cite{Nakahara} we obtain
\bsubeqs 	\label{eq:cappuccio_first}
\begin{equation}  \label{eq:eta_munu_cappuccio}
\boldsymbol{g}  =\eta_{\hat{\mu}\hat{\nu}} \, \boldsymbol{\theta}^{\hat{\mu}} \otimes \boldsymbol{\theta}^{\hat{\nu}},
\end{equation}
\begin{equation} \label{eq:theta_t}
\boldsymbol{\theta}^{\hat{\tau}}= \sqrt{\dfrac{t^2}{t^2+b^2}}\,\bold{d}t,
\end{equation}
\begin{equation}
\boldsymbol{\theta}^{\hat{x}^i} = a(t) \,\bold{d}x^i.
\end{equation}
\esubeqs
From the relations
\bsubeqs
\begin{equation}
\boldsymbol{\theta}^{\hat{\alpha}} = e^{\hat{\alpha}}_{\phantom{\hat{\alpha}} \mu}\, \bold{d}x^\mu,
\end{equation}
\begin{equation}
e^{\hat{\alpha}}_{\phantom{\hat{\alpha}} \mu} e_{\hat{\beta}}^{{\phantom{\hat{\beta}} \mu}}= \delta^{\hat{\alpha}}_{\hat{\beta}},
\end{equation}
\begin{equation}
\bold{e}_{\hat{\alpha}}= e_{\hat{\alpha}}^{{\phantom{\hat{\alpha}} \mu}} \bold{e}_\mu,
\end{equation}
\esubeqs
we get (see Eq. (\ref{eq:Eulerian_velocity}))
\bsubeqs \label{eq: e_alpha_cappuccio}
\begin{equation}\label{eq:e_0 and n}
\bold{e}_{\hat{\tau}} = \sqrt{\dfrac{t^2+b^2}{t^2}}   \, \bold{e}_{t} = \boldsymbol{n},
\end{equation}
\begin{equation}
\bold{e}_{\hat{x}^i} = \dfrac{1}{a(t)} \bold{e}_{x^i}.
\end{equation}
\esubeqs
The components $R_{\hat{\alpha}\hat{\beta}\hat{\gamma}\hat{\delta}}$ of the Riemann tensor in the proper reference frame $(\bold{e}_{\hat{\tau}},\bold{e}_{\hat{x}},\bold{e}_{\hat{y}},\bold{e}_{\hat{z}})$ can be obtained from those computed in the frame $\bold{e}_\mu=(\bold{e}_t,\bold{e}_x,\bold{e}_y,\bold{e}_z)$ through
\bsubeqs \label{eq:Riemann_hat_indices_def}
\begin{equation}
R_{\hat{\alpha}\hat{\beta}\hat{\gamma}\hat{\delta}} = e_{\hat{\alpha}}^{\phantom{\hat{\alpha}}\alpha}e_{\hat{\beta}}^{\phantom{\hat{\beta}}\beta}e_{\hat{\gamma}}^{\phantom{\hat{\gamma}}\gamma}e_{\hat{\delta}}^{\phantom{\hat{\delta}}\delta} R_{\alpha \beta \gamma \delta},
\end{equation}
\begin{equation}
R^{\hat{\alpha}}_{ \phantom{\hat{\alpha}}\hat{\beta}\hat{\gamma}\hat{\delta}} = \eta^{\hat{\alpha}\hat{\mu}} R_{\hat{\mu}\hat{\beta}\hat{\gamma}\hat{\delta}}. 
\end{equation}
\esubeqs

At this stage, we can analyze the compressive forces acting on the Eulerian human observer by evaluating the geodesic deviation equation in his/her proper reference. If  we set $\boldsymbol{u}=\boldsymbol{n}$ in Eq.  (\ref{eq:geodesic_deviation_equation}) and exploit  Eqs.  (\ref{eq:e_0 and n}) and (\ref{eq:Riemann_hat_indices_def}) along with the condition $\boldsymbol{\xi}\cdot \boldsymbol{n}=-\xi^{\hat{\tau}}=0$, we obtain (no sum over $\hat{j}$):
\bsubeqs \label{eq:compressive_on_Eulerian}
\begin{equation}
\Delta a^{\hat{j}} =-R^{\hat{j}}_{\phantom{\hat{j}}\hat{\tau}\hat{j}\hat{\tau}} \,  \xi^{\hat{j}},
\end{equation}
\begin{equation}
\Delta a^{\hat{j}} = \dfrac{{\rm d}^2}{{\rm d}\tau^{2}} \, \xi^{\hat{j}},
\end{equation}
\begin{equation}\label{eq:Riemann_hat_indices}
-R^{\hat{j}}_{\phantom{\hat{j}}\hat{\tau}\hat{j}\hat{\tau}}  =\left[\left(\dfrac{b^2+t^2}{t^2}\right)\dfrac{\ddot{a}}{a}+\left(-\dfrac{b^2}{t^3}\,\dfrac{\dot{a}}{a}\right)\right] \\
 =\left \{ \begin{array}{rl} 
& -\dfrac{2 }{9\left(b^2+t^2\right)}, \quad {\rm nonrelativistic\; matter},\\
&  -\dfrac{1}{4\left(b^2+t^2\right)}, \quad {\rm relativistic\; matter}.
\end{array} \right.
\end{equation}
\esubeqs
The fact that Eq. (\ref{eq:compressive_on_Eulerian}) depends on both first order and second order derivatives of $a(t)$ marks a difference with standard RW cosmology, which predicts compressive forces depending only on the ratio $\ddot{a}/a$. On the other hand, the minus sign occurring in front of the quantities contained inside the curly bracket in Eq. (\ref{eq:Riemann_hat_indices})  shows that we are dealing with  compressive forces (cf. Eq. (\ref{eq:geodesic_deviation_equation})), in strict analogy with standard RW cosmology \cite{Wald}, where Friedmann equations foretell a scale factor having $\ddot{a}<0$ as long as $\rho >0$ and  $P \geq 0$ ($\rho$ and $P$ being, as pointed out in Sec. \ref{Sec:Freely_falling_obs_FLRW}, the matter energy density and the matter pressure, respectively). 

Equation (\ref{eq:Riemann_hat_indices}) implies that compressive forces  acting at $t=0$ on the Eulerian human observer are given by 
\begin{equation} \label{eq:tidal_forces_Eulerian_2}
\begin{split}
&\lim_{t\rightarrow 0} \left(-R_{\hat{x}\hat{\tau}\hat{x}\hat{\tau}}\right)=\lim_{t\rightarrow 0} \left(-R_{\hat{y}\hat{\tau}\hat{y}\hat{\tau}}\right)=\lim_{t\rightarrow 0} \left(-R_{\hat{z}\hat{\tau}\hat{z}\hat{\tau}}\right) \\
& =\left \{ \begin{array}{rl} 
& -2/(9b^2), \quad {\rm nonrelativistic\; matter},\\
& -1/(4b^2), \quad {\rm relativistic\; matter},
\end{array} \right.
\end{split}
\end{equation}
meaning that  such observer is subjected to  forces  inversely proportional to the square of the length scale $b$ when he/she gets at the defect.

Let us analyze separately the two terms occurring in the square brackets of  Eq. (\ref{eq:Riemann_hat_indices}). First of all, we have
\begin{equation} 
\label{eq:aduepunti/a_1}
\begin{split}
\dfrac{\ddot{a}(t)}{a(t)}
& =\left \{ 
\setlength{\tabcolsep}{10pt} 
\renewcommand{\arraystretch}{1.8}
\begin{array}{rl} 
& \dfrac{2 \left(3b^2-t^2\right)}{9\left(b^2+t^2\right)^2}, \quad {\rm nonrelativistic\; matter},\\
& \dfrac{ \left(2b^2-t^2\right)}{4\left(b^2+t^2\right)^2}, \quad \,{\rm relativistic\; matter},
\end{array} \right.
\end{split}
\end{equation}
which means that in a finite time interval around  $t=0$ (i.e., near the defect) we have\footnote{In Eq. (\ref{eq:Riemann_hat_indices})  the factor $\ddot{a}/a$ is multiplied by the term $(b^2+t^2)/t^2$ which  is always positive except at $t=0$, where it is not defined.} 
\begin{equation} 
\label{eq:aduepunti/a_2}
\begin{split}
\dfrac{\ddot{a}(t)}{a(t)} >0 \Leftrightarrow
& \left \{ \begin{array}{rl} 
& -\sqrt{3}\,b<t<\sqrt{3}\,b, \quad \;{\rm nonrelativistic\; matter},\\
& -\sqrt{2}\,b<t<\sqrt{2}\,b, \quad \;{\rm relativistic\; matter}.
\end{array} \right.
\end{split}
\end{equation}
Equations (\ref{eq:aduepunti/a_1}) and (\ref{eq:aduepunti/a_2}) imply that the factor $\ddot{a}/a$ occurring in  Eq. (\ref{eq:Riemann_hat_indices}) produces stretching forces in the neighbourhood of the defect  and compressive forces  far from it. In particular, if we define 
\bsubeqs 	\label{eq:t_star}
\begin{equation}
t^{\star} \equiv \sqrt{n}\, b,
\end{equation} 
\begin{equation}
\begin{split}
n=
& \left \{ \begin{array}{rl} 
& 3, \quad {\rm nonrelativistic\; matter},\\
& 2, \quad {\rm relativistic\; matter},
\end{array} \right.
\end{split}
\end{equation}
\esubeqs
we can say that  if $t<-t^{\star}$ (contracting-universe phase) or $t>t^{\star}$ (expanding-universe phase) the term  $\ddot{a}/a$ generates a compressive force, whereas if $-t^{\star} <t<0$ (contracting epoch) or $0<t<t^{\star} $ (expanding epoch)  $\ddot{a}/a$ leads to stretching forces. Here, we can appreciate the antigravitational action of the defect, which produces an accelerated  contraction or expansion rate of the universe for $t\in \left(-t^{\star},t^{\star}\right)$.

The second term appearing in the square brackets of  Eq. (\ref{eq:Riemann_hat_indices}) always causes compressive forces, since we have
\begin{equation} 
\begin{split}
-\dfrac{\dot{a}(t)}{a(t)} \dfrac{b^2}{t^3}
& =\left \{ 
\setlength{\tabcolsep}{10pt} 
\renewcommand{\arraystretch}{1.8}
\begin{array}{rl} 
& -\dfrac{2 b^2}{3\left(b^2+t^2\right)}, \quad {\rm nonrelativistic\; matter},\\
&  -\dfrac{b^2}{2\left(b^2+t^2\right)}, \quad {\rm relativistic\; matter}.
\end{array} \right.
\end{split}
\end{equation}
However, we have already seen that  the sum of the two contributions of Eq. (\ref{eq:Riemann_hat_indices})  amounts to a compressive force. This means that near the defect, where the terms $\ddot{a}/a$ and $[-(\dot{a}/a)(b^2/t^3)]$ give opposite contributions, the modulus of $[-(\dot{a}/a)(b^2/t^3)]$ \qm{wins} against $\ddot{a}/a$ (which is positive for $-t^\star<t<t^\star$, cf. Eq. (\ref{eq:aduepunti/a_2})) so that their sum gives a compressive force.

In principle, it could be possible to build a scenario where both compressive and stretching forces act on the Eulerian human observer in the vicinity of the defect. However, this would require, in the  spacetime (\ref{modified_FRW}), a  scale factor  for which (at least) third-order time derivatives are  nonvanishing at $t=0$. In other words, such a model would entail an odd $a(t)$ function.

For future purposes, it is important to express the components $R^{\hat{j}}_{\phantom{\hat{j}}\hat{\tau}\hat{j}\hat{\tau}} $ (no sum over $\hat{j}$) of the Riemann tensor appearing in Eq. (\ref{eq:compressive_on_Eulerian})  in an equivalent  way. Indeed, the Riemann tensor written in terms of  the Christoffel symbols $\Gamma^{\hat{\gamma}}_{\phantom{\hat{\gamma}}\hat{\alpha}\hat{\beta}}$ only reads as \cite{Nakahara}
\begin{equation} \label{eq:Riemann_hat_indices_2}
R^{\hat{\alpha}}_{\phantom{\hat{\alpha}}\hat{\beta}\hat{\gamma}\hat{\delta}}= \bold{e}_{\hat{\gamma}} \left[ \Gamma^{\hat{\alpha}}_{\phantom{\hat{\alpha}}\hat{\delta}\hat{\beta}} \right]-\bold{e}_{\hat{\delta}} \left[ \Gamma^{\hat{\alpha}}_{\phantom{\hat{\alpha}}\hat{\gamma}\hat{\beta}} \right]+\Gamma^{\hat{\epsilon}}_{\phantom{\hat{\epsilon}}\hat{\delta}\hat{\beta}}\Gamma^{\hat{\alpha}}_{\phantom{\hat{\alpha}}\hat{\gamma}\hat{\epsilon}}-\Gamma^{\hat{\epsilon}}_{\phantom{\hat{\epsilon}}\hat{\gamma}\hat{\beta}}\Gamma^{\hat{\alpha}}_{\phantom{\hat{\alpha}}\hat{\delta}\hat{\epsilon}} - \Gamma^{\hat{\alpha}}_{\phantom{\hat{\alpha}}\hat{\epsilon}\hat{\beta}} \left( \Gamma^{\hat{\epsilon}}_{\phantom{\hat{\epsilon}}\hat{\gamma}\hat{\delta}} -\Gamma^{\hat{\epsilon}}_{\phantom{\hat{\epsilon}}\hat{\delta}\hat{\gamma}}\right),
\end{equation}
and the relation between the connection coefficients $\Gamma^{\hat{\gamma}}_{\phantom{\hat{\gamma}}\hat{\alpha}\hat{\beta}}$ in the basis $\{\bold{e}_{\hat{\alpha}}\}$ and $\Gamma^{\nu}_{\phantom{\nu}\mu \lambda}$ in the basis $\{\bold{e}_{\mu}\}$ is given by \cite{Nakahara}
\begin{equation} \label{eq:gamma_hat_indeces}
\Gamma^{\hat{\gamma}}_{\phantom{\hat{\gamma}}\hat{\alpha}\hat{\beta}}= e^{\hat{\gamma}}_{\phantom{\hat{\gamma}}\nu} e_{\hat{\alpha}}^{\phantom{\hat{\alpha}}\mu} \left( \partial_\mu e_{\hat{\beta}}^{\phantom{\hat{\beta}}\nu}  + e_{\hat{\beta}}^{\phantom{\hat{\beta}}\lambda}\,  \Gamma^{\nu}_{\phantom{\nu}\mu \lambda} \right).
\end{equation}
Therefore, from Eqs. (\ref{eq:Riemann_hat_indices_2}) and (\ref{eq:gamma_hat_indeces})  we find
\begin{equation} \label{eq:Riemann_hat_indices_3}
 -R^{\hat{j}}_{\phantom{\hat{j}}\hat{\tau}\hat{j}\hat{\tau}}  = \bold{e}_{\hat{\tau}}  \left[ \Gamma^{\hat{j}}_{\phantom{\hat{j}}\hat{j}\hat{\tau}} \right] + \left(\Gamma^{\hat{j}}_{\phantom{\hat{j}}\hat{j}\hat{\tau}}\right)^2 = e_{\hat{\tau}}^{\phantom{\hat{\tau}}t} \, \bold{e}_t  \left[e_{\hat{\tau}}^{\phantom{\hat{\tau}}t} \, \Gamma^{j}_{\phantom{j}jt} \right] + \left(e_{\hat{\tau}}^{\phantom{\hat{\tau}}t} \, \Gamma^{j}_{\phantom{j}jt} \right)^2, \quad ({\rm no\; sum\; over}\; \hat{j},j),
\end{equation} 
which, after having exploited Eqs. (\ref{eq:cappuccio_first})--(\ref{eq: e_alpha_cappuccio}), leads to the same result as Eq. (\ref{eq:Riemann_hat_indices}).

The most important aspect of the above calculation is that  Eq. (\ref{eq:Riemann_hat_indices_3}) has no contribution from $\Gamma^{t}_{\phantom{t}tt}$ (i.e., the only Christoffel symbol of (\ref{modified_FRW}) which does not depend on $a(t)$ and its derivatives, see Eq. (\ref{eq:Gamma_sec_2})). Indeed, this observation will be crucial in Sec. \ref{Sec:Energy}, where we will claim that  compressive forces (\ref{eq:compressive_on_Eulerian}) show no substantial differences with respect to the corresponding standard cosmology case because  the Eulerian observer cannot measure  contributions related to $\Gamma^{t}_{\phantom{t}tt}$ in his/her proper reference frame (see the discussion below Eq. (\ref{eq:Gamma^0_mu-nu cappuccio})).

At this stage, we are ready to move to the proper reference frame $(\bold{e}_{\hat{\tau}^{\prime}},\bold{e}_{\hat{x}^{\prime}},\bold{e}_{\hat{y}^{\prime}},\bold{e}_{\hat{z}^{\prime}})$ of the traveller by means of a boost along the $\bold{e}_{\hat{x}}$ direction. We can calculate the ordinary boost velocity $V^{\hat{x}}$ as
\begin{equation} \label{eq:boost_velocity}
\begin{split}
V^{\hat{x}} &= \dfrac{{\rm proper\, distance\, along\, }\bold{e}_{\hat{x}}\, {\rm as\, seen\, by\, the\, Eulerian\, observer}}{{\rm proper\, lapse\, time \,as \,seen\, by \,the \,Eulerian\, observer}} \\
& =\dfrac{\sqrt{g_{xx}}\,{\rm d}x}{\sqrt{-g_{tt}}\,{\rm d}t}= \dfrac{1}{\sqrt{1+a(t)^2/\Pi^2}} \equiv V,
\end{split}
\end{equation} 
where we have exploited Eq. (\ref{eq:traveler_geodesic}). In terms of the orthonormal bases $\{\bold{e}_{\hat{\alpha}}\}$ of the Eulerian observer and $\{\bold{e}_{\hat{\alpha}^\prime}\}$ of the traveller, the boost is expressed by
\bsubeqs \label{eq:e-alpha_cappuccio_prime}
\begin{equation} 
\left \{ \begin{array}{rl} 
& \bold{e}_{\hat{\tau}^\prime}=\gamma \bold{e}_{\hat{\tau}} + \gamma V\bold{e}_{\hat{x}} ,\\
& \bold{e}_{\hat{x}^\prime}=\gamma V \bold{e}_{\hat{\tau}} + \gamma \bold{e}_{\hat{x}} ,\\
& \bold{e}_{\hat{y}^\prime}= \bold{e}_{\hat{y}},\\
& \bold{e}_{\hat{z}^\prime}= \bold{e}_{\hat{z}},
\end{array} \right.
\end{equation}
\begin{equation}
\bold{e}_{\hat{\tau}^\prime} = \boldsymbol{v},
\end{equation}
\begin{equation} \label{eq:gamma_factor}
\gamma=\dfrac{1}{\sqrt{1-V^2}}= \sqrt{1+\Pi^2/a(t)^2}.
\end{equation}
\esubeqs
The components $R_{\hat{\alpha}^\prime\hat{\beta}^\prime\hat{\gamma}^\prime\hat{\delta}^\prime}$ of the Riemann tensor  in the traveller's proper frame can be obtained by applying the usual Lorentz transformation to  the components $R_{\hat{\alpha}\hat{\beta}\hat{\gamma}\hat{\delta}}$ measured by the Eulerian observer in his/her proper frame. Moreover, since the separation vector $\boldsymbol{\xi}_T$ is purely spatial in the traveller's frame,  the geodesic deviation equation (\ref{eq:geodesic_deviation_equation}) can be written as
\bsubeqs \label{eq:tidal_traveller_3}
\begin{equation}
\Delta a^{\hat{i}^\prime}=\Delta a_{\hat{i}^\prime}= -R_{\hat{i}^\prime\hat{\tau}^\prime\hat{j}^\prime\hat{\tau}^\prime} \, \xi_T^{\hat{j}^\prime}, 
\end{equation}
\begin{equation}
\Delta a^{\hat{i}^\prime} = \dfrac{{\rm d}^2}{{\rm d}\tau^{\prime\,2}} \, \xi_T^{\hat{i}^\prime}.
\end{equation}
\esubeqs
Bearing in mind the above equation, the compressive forces felt by the human traveller as he/she passes through the defect are given by
\bsubeqs \label{eq:tidal_traveller}
\begin{equation} 
\lim_{t \rightarrow 0} \left(-R_{\hat{x}^\prime\hat{\tau}^\prime\hat{x}^\prime\hat{\tau}^\prime} \right) =\left \{ \begin{array}{rl} 
& -2/(9b^2), \quad {\rm nonrelativistic\; matter},\\
& -1/(4b^2), \quad {\rm relativistic\; matter},
\end{array} \right.
\end{equation}
\begin{equation} \label{eq:tidal_traveller_2}
\begin{split}
& \lim_{t \rightarrow 0} \left(-R_{\hat{y}^\prime\hat{\tau}^\prime\hat{y}^\prime\hat{\tau}^\prime} \right) =\lim_{t \rightarrow 0} \left(-R_{\hat{z}^\prime\hat{\tau}^\prime\hat{z}^\prime\hat{\tau}^\prime} \right) =\\ 
&\left \{
\begin{array}{rl} 
& -[2/(9b^2)](1+3 \Pi^2/a(0)^2), \quad \quad{\rm nonrelativistic\; matter},\\
& -[1/(4b^2)] (1+2 \Pi^2/a(0)^2), \quad \quad {\rm relativistic\; matter}.
\end{array} \right.
\end{split}
\end{equation}
\esubeqs 
At this stage, if we take into account the hypothesis  developed in Ref. \cite{KII} according to which the defect length scale $b$ can be of order of the Planck length ${\ell}_{\rm P}$ and we also suppose that
\bsubeqs \label{eq:hypothesis}
\begin{equation} \label{eq:hypothesis_1}
 t_0 \gg b,
\end{equation}
\begin{equation}
 \dfrac{ \vert\Pi  \vert t_0}{b} \gg 1,
\end{equation}
\esubeqs
we can write (\ref{eq:tidal_traveller_2}) approximately as
\begin{equation} \label{eq:tidal_traveller_approx}
\begin{split}
& \lim_{t \rightarrow 0} \left(-R_{\hat{y}^\prime\hat{\tau}^\prime\hat{y}^\prime\hat{\tau}^\prime} \right) =\lim_{t \rightarrow 0} \left(-R_{\hat{z}^\prime\hat{\tau}^\prime\hat{z}^\prime\hat{\tau}^\prime} \right) \approx \\
& \left \{
\begin{array}{rl} 
& -[2 \Pi^2 (t_0)^{4/3}]/[3b^2 (b)^{4/3}], \quad \;{\rm nonrelativistic\; matter},\\
& -[ \Pi^2 \, \vert t_0\vert\, ]/[2b^3], \quad \quad \quad \quad \quad \,{\rm relativistic\; matter}.
\end{array} \right.
\end{split}
\end{equation}

\begin{figure}[th!]
\centering
\includegraphics[scale=0.74]{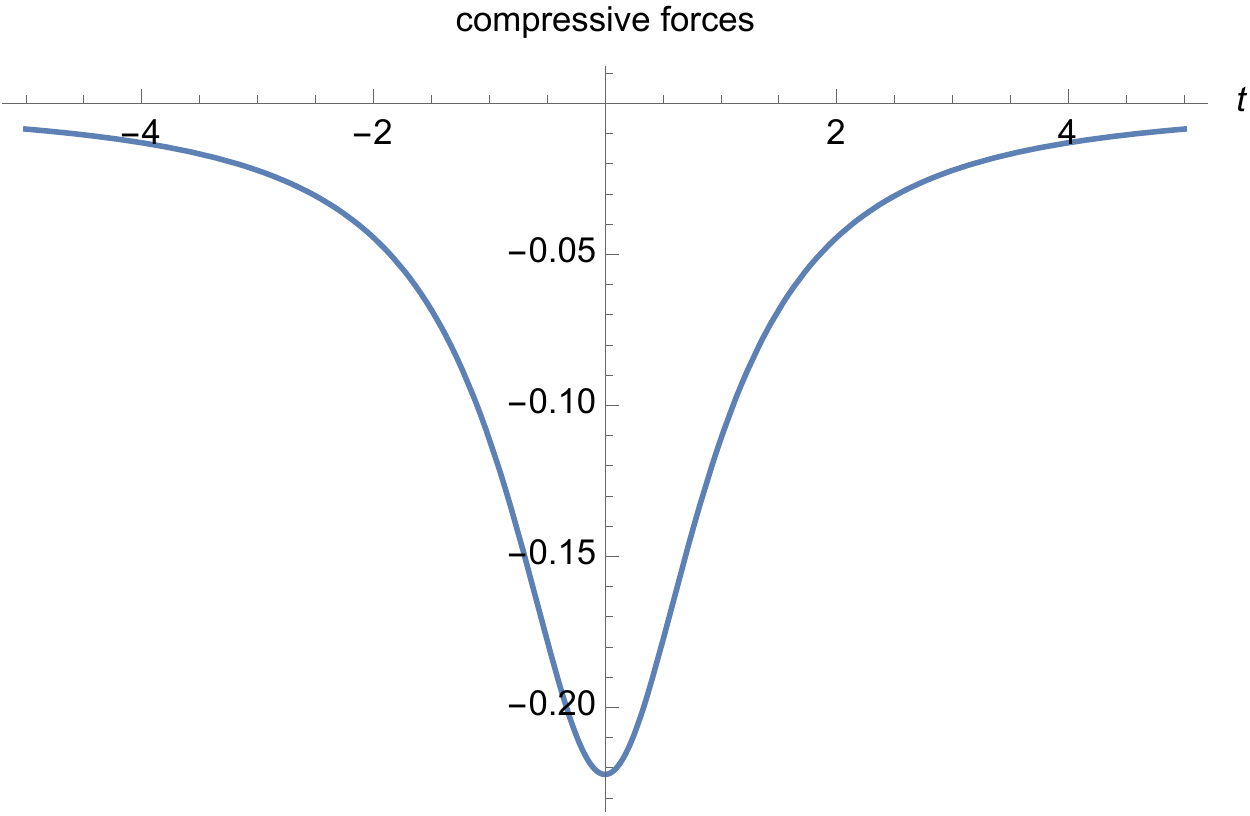}
\caption{Compressive forces acting on the Eulerian observer for the nonrelativistic-matter solution (\ref{eq:scalefactor}) and with $b=1$ (cf. Eq. (\ref{eq:Riemann_hat_indices})). A similar plot is obtained for compressive forces experienced by the traveller.}
\label{fig:compressive_forces}
\end{figure}

Some comments on the results obtained so far are in order. First of all, Eqs. (\ref{eq:tidal_forces_Eulerian_2}) and  (\ref{eq:tidal_traveller}) show that compressive forces are finite at $t=0$ provided that $b\neq 0$. This may be seen as   an equivalent proof of the fact  that  in the RW geometry (\ref{modified_FRW})  the big-bang singularity can be regularized  via the introduction of the defect \cite{KI}. Furthermore, the analysis of compressive forces acting on both the Eulerian observer and the traveller reveals that the spacetime defect can be  identified as the three-dimensional spacelike hypersurface of the  regularized RW spacetime  where compressive forces become as intense as possible. This is clear from Fig. \ref{fig:compressive_forces}, where we have chosen to plot compressive forces experienced by the Eulerian observer in the case of nonrelativistic-matter solution (cf. Eq. (\ref{eq:scalefactor})). Compressive forces on the Eulerian observer for the relativistic-matter solution and on the traveller (for both the matter-dominated and the radiation-dominated universe) display the same behaviour as the one in Fig. \ref{fig:compressive_forces}.
Moreover, Eqs. (\ref{eq:tidal_forces_Eulerian_2}), (\ref{eq:tidal_traveller}) and (\ref{eq:tidal_traveller_approx}) show that if $b \sim {\ell}_{\rm P}$ then both the Eulerian human observer and the human traveller are subjected to very large compressive forces proportional to $1/b$ as they pass the defect, which therefore amounts to a gravitational obstacle between two different universes, the one with $t <0$ and that with $t>0$. 
To have an idea, a rough calculation reveals that if we assume that a human body cannot withstand an acceleration gradient of ten times the Earth's gravity (i.e., $10g_{\oplus} \approx 98$ m/s$^2$) per metre, then Eq. (\ref{eq:tidal_forces_Eulerian_2}) implies that the Eulerian human observer  can survive the compressive forces generated at $t=0$ only if $b \gtrsim 10^7$ m.  This suggests that the results of this section can concur  with the analysis performed in Refs. \cite{KII,KWIII,K2020-IIB-3} (see also Ref. \cite{Klinkhamer2021a}) where  a quantum origin of the defect length scale is proposed.  Indeed, as spelled out in  Ref. \cite{Klinkhamer2021a},  our model (\ref{modified_FRW}) could yield  two possible scenarios, depending on the values assumed by the defect length scale $b$. The first pattern leads to a nonsingular-bouncing-cosmology model, whereas the second provides  a new physics phase at $t=0$ which  pair-produces  a \qm{universe} for $t>0$ and an \qm{anti-universe} for $t<0$. The first framework is built on classical Einstein theory and hence may be possible  if $b \gg {\ell}_{\rm P}$, whereas the second might apply if $b \sim {\ell}_{\rm P}$. Our investigation shares 
some similar features with this last situation since we have shown that  if $b \sim {\ell}_{\rm P}$ then compressive forces become so large at $t=0$ that the universe having $t<0$  and  the one with $t>0$ can be viewed as separated.

It should be stressed that Eqs. (\ref{eq:compressive_on_Eulerian}) and (\ref{eq:tidal_traveller_3}) (likewise those derivable thereof) will not hold exactly, since they do not account for the forces between the  atoms comprising the observers. Nevertheless, when gravitational compressive forces become very strong  such interatomic forces can be neglected and hence the equations derived above will be valid to  good approximation. Therefore, we can conclude that when compressive forces tear the Eulerian human observer and the human traveller apart,  the very atoms of which they are  composed must ultimately undergo the same fate.

\section{The energy of massive particles and photons} \label{Sec:Energy}

We have examined, in the previous section, compressive forces acting on the  Eulerian human observer and the human traveller. In order to  complete our physical description  of the spacetime defect underlying the modified RW cosmology (\ref{modified_FRW}),  we will now analyze the energy behaviour of  both the Eulerian observer and the traveller during their motion. Furthermore, our investigation will involve  the energy of  photons. We also note that for these considerations we have no need to think of the Eulerian observer and the traveller as human observers.

First of all, we need to define the concept of massive particle or photon energy in a proper way. Before we set out this theme, let us describe two simple situations where  we can safely define this physical quantity. As a first example,   consider the Schwarzschild solution. In  this  case, it is possible to define the \emph{total} energy of a massive particle or a photon due to the presence of a timelike Killing vector field related to the time-translation invariance of Schwarzschild geometry; moreover, in the case of timelike geodesics, this conserved quantity reduces, at large distances from the center of attraction, to the usual special relativistic formula for the total energy  per unit mass of the particle \cite{Wald}. Another example is furnished by the energy of a massive particle or a photon with four-momentum $\boldsymbol{p}$ which is (locally) measured by a generic observer having four-velocity $\boldsymbol{u}_{\rm obs}$  in his/her proper reference frame (whose orthonormal basis four-vectors are such that $\bold{e}_{\hat{0}}=\boldsymbol{u}_{\rm obs}$\footnote{We have labelled the orthonormal basis vector  as $\bold{e}_{\hat{0}}$ and not as $\bold{e}_{\hat{\tau}}$  or $\bold{e}_{\hat{\tau}^{\prime}}$ because the measurement process (\ref{eq:E_measured}) can be performed by any observer, not only those characterized by a free-fall motion.}).  In this case, we find
\begin{equation} \label{eq:E_measured}
E_{\rm measured}= - \boldsymbol{p} \cdot \boldsymbol{u}_{\rm obs} = p^{\hat{0}} =\langle\boldsymbol{\theta}^{\hat{0}}, \boldsymbol{p}\rangle,
\end{equation}
where $\{\boldsymbol{\theta}^{\hat{\alpha}}\}$ is the basis one-forms of the observer's proper reference frame (cf. Eq. (\ref{eq:eta_munu_cappuccio})) and $\langle \cdot \, ,\, \cdot \rangle$ is the usual inner product. However, in this case thre is a caveat:  expression (\ref{eq:E_measured}) only represents an \emph{intrinsic} energy from motion and inertia, \emph{not} the \emph{total} energy of the massive particle or the photon.  In that regard, we will see that in our model Eq. (\ref{eq:E_measured}) yields, in agreement with the equivalence principle (which holds for all times except at $t= 0$),  an expression for the energy measured by the Eulerian observer where there is no room for the term $\Gamma^{t}_{\phantom{t}tt}$, which, as  we have pointed out in Sec. \ref{Sec:Compressive_forces}, fulfils an important role in our analysis. Indeed, we will interpret this $\Gamma^{t}_{\phantom{t}tt}$  term as due to the (anti-)gravitational action of the defect (see comments above Eq. (\ref{eq:expression_of_Gamma_ttt})).

To formulate an energy definition suitable for our model, we need to briefly recall some topics. It is well-known  that the Hamiltonian formulation of general relativity as well as the analysis of the Cauchy (initial value) problem can be performed by employing the  $3+1$ formalism, which relies on a theorem stating  that any globally hyperbolic spacetime can be foliated by a constant-$t$ family $(\Sigma_t)_{t \in \mathbb{R}}$ of spacelike (Cauchy) hypersurfaces \cite{Wald,Gourgoulhon,Choquet-Bruhat2014,Hawking-Ellis}.  Within this framework, the spacetime metric can be written as 
\begin{equation} \label{eq:metric,lapse,shift}
g_{\mu \nu} = \begin{bmatrix}
N_i N^i -N^2 & N_k \\
N_k & h_{ik}
\end{bmatrix},
\end{equation}
where $N$, $N_i$  are the lapse function and the shift vector, respectively, while $h_{ik}$ is the (induced) 3-metric on the generic hypersurface $\Sigma_t$ belonging to the set $(\Sigma_t)_{t \in \mathbb{R}}$. An inspection of Eq. (\ref{modified_line_element}), reveals that in our model we have
\bsubeqs \label{foliation}
\begin{equation} \label{lapse_function}
N= \dfrac{\vert t \vert }{\sqrt{t^2+b^2}},
\end{equation}
\begin{equation} \label{shift_vector}
N_i=0,
\end{equation}
\begin{equation}
h_{ik}= a(t)^2 \delta_{ik}.
\end{equation}
\esubeqs
Furthermore, in the context of the geometry of foliation the Eulerian observer four-velocity can be written in general as $n^\alpha= (\dfrac{1}{N},\dfrac{-N^i}{N})$ and it represents the timelike and future-oriented unit four-vector normal to the generic hypersurface $\Sigma_t$. In our modified RW setup, it follows from Eqs. (\ref{lapse_function}) and (\ref{shift_vector})  that Eq. (\ref{eq:Eulerian_velocity}) can be written equivalently as
\begin{equation} \label{eq:Eulerian_velocity_with_lapse}
n^\alpha= \dfrac{1}{N}(1,0,0,0).
\end{equation}
An important object of the $3+1$ approach is the so-called  normal evolution (four-)vector \cite{Gourgoulhon}, defined as the timelike vector field
\begin{equation} \label{eq:normal_evolution_vector}
\boldsymbol{m} \equiv N \boldsymbol{n},
\end{equation}
which carries (or Lie drags) the hypersurface $\Sigma_t$ (defined, as pointed out before, by the condition $t={\rm const}$) to the neighbouring hypersurface $\Sigma_{t+{\rm d}t}$. In other words, the hypersurface $\Sigma_{t+{\rm d}t}$ can be obtained from the neighbouring $\Sigma_t$ by a small displacement $\boldsymbol{m}\,{\rm d}t$ of each point of $\Sigma_t$. From Eq. (\ref{eq:Eulerian_velocity_with_lapse}) we easily find that  our regularized RW geometry is characterized by the following normal evolution (four-)vector:
\begin{equation} \label{eq:modified_normal_evolution_vector}
\boldsymbol{m}=\bold{e}_t.
\end{equation}

Bearing in mind the above results, we propose for  model (\ref{modified_FRW}) the following definition of the \emph{total} energy. We identify  the total energy $E_{\rm tot}$ of  a massive particle or a photon having four-momentum $\boldsymbol{p}$ with the projection of  $\boldsymbol{p}$ along the normal evolution (four-)vector, i.e.,
\begin{equation} \label{eq:Energy_tot}
E_{\rm tot} = - \boldsymbol{p} \cdot \boldsymbol{m}.
\end{equation}
This is justified by the fact that, as explained before, the four-vector $\boldsymbol{m}$ regulates the temporal evolution of the spacetime (note however that this evolution is described in terms of the coordinate time variable $t$, see below). Therefore, the procedure underlying Eq. (\ref{eq:Energy_tot}) turns out to be similar to the method  adopted by an observer who measures the (intrinsic) energy of a massive particle/photon in his/her proper reference frame by simply  projecting the massive particle/photon four-momentum $\boldsymbol{p}$ along $\bold{e}_{\hat{0}}=\boldsymbol{u}_{\rm obs}$, i.e., along the time direction defined by the observer's clock (cf. Eq. (\ref{eq:E_measured})). In a similar way, we propose through Eq. (\ref{eq:Energy_tot}) to define the total energy by projecting  $\boldsymbol{p}$ along the four-vector $\boldsymbol{m}$ ruling the evolution of the spacetime.  We will justify later in which sense  (\ref{eq:Energy_tot}) defines a \emph{total} energy (see comments below Eq. (\ref{eq:expression_of_Gamma_ttt})). However, in our definition (\ref{eq:Energy_tot}) there is one important aspect which must be taken into   account: while $\bold{e}_{\hat{0}}$ governs the flow of the observer's proper time,  $\boldsymbol{m}$ can be seen as the vector controlling the flow of the coordinate time.

We can analyze the consequences of our definition (\ref{eq:Energy_tot}) by considering the form assumed by $E_{\rm tot}$ in the most general situations. Consider for instance the generic four-dimensional spacetime metric written as in Eq. (\ref{eq:metric,lapse,shift}). According to our definition (\ref{eq:Energy_tot}), the total energy $\mathcal{E}$ of the Eulerian observer is given by (for the sake of simplicity,  the rest masses of all observers are set to one)
\begin{equation} \label{eq:energy_eulerian}
\mathcal{E} = - \boldsymbol{n} \cdot \boldsymbol{m} = N = -n_{t}.
\end{equation} 
Now consider a generic particle having four-momentum $\boldsymbol{p}$. According to our prescription (\ref{eq:Energy_tot}), the total energy will be
\begin{equation} \label{eq:e_particle_1}
E_{\rm particle}= -\boldsymbol{p} \cdot \boldsymbol{m}= N^2 p^t.
\end{equation}
In this case, $E_{\rm particle} \neq -p_t$ since
\begin{equation} \label{eq:e_particle_2}
-p_t = N^2 p^t -N_iN^i p^t -N_i p^i,
\end{equation} 
and hence 
\begin{equation} \label{eq:e_particle_3}
E_{\rm particle} =  -p_t \Leftrightarrow N^i=0.
\end{equation}
Equations (\ref{eq:e_particle_1})--(\ref{eq:e_particle_3}) reveal the advantages of our definition (\ref{eq:Energy_tot}). Indeed, Eq. (\ref{eq:e_particle_1}) depends only on the gauge function $N$ (reflecting thus the fact that the energy is not a scalar and hence its expression changes according to the coordinates adopted), whereas (\ref{eq:e_particle_2}) depends on both $N$ and $N^i$. However, this last circumstance would lead to an ill-defined concept of energy since the contributions  due to $N^i$ can always  be gauged away once comoving spatial coordinates are invoked. On the contrary, by adopting the definition (\ref{eq:e_particle_1}), the energy does not depend on terms which can be set to zero by an appropriate coordinate transformation. 

In  model (\ref{modified_FRW}), the normal evolution (four-)vector is represented by Eq. (\ref{eq:modified_normal_evolution_vector}) and hence we can define the \emph{total} energy of the traveller and the photon as, respectively,
\bsubeqs
\begin{equation}
E=  -\boldsymbol{v} \cdot \boldsymbol{m} = -v_t,
\end{equation}
\begin{equation}
E_{\rm ph} = -\boldsymbol{k} \cdot \boldsymbol{m} = -k_t,
\end{equation}
\esubeqs
$\boldsymbol{k} $ being the photon four momentum.  In addition, the total energy of the Eulerian observer will be represented by (\ref{eq:energy_eulerian}), which by means of  Eq. (\ref{lapse_function})  reads as
\begin{equation} \label{eq:Eulerian_energy}
\mathcal{E}= \sqrt{\dfrac{t^2}{t^2+b^2}}.
\end{equation}

Let $E_{\rm local}$ represent the traveller energy as measured by the Eulerian observer in his/her proper reference frame   $(\bold{e}_{\hat{\tau}},\bold{e}_{\hat{x}},\bold{e}_{\hat{y}},\bold{e}_{\hat{z}})$.  Then we have from Eq. (\ref{eq:E_measured})
\begin{equation} \label{eq:E_local}
E_{\rm local} = -\boldsymbol{v}\cdot \boldsymbol{n}=- v_{\hat{\tau}}=v^{\hat{\tau}} =\langle \boldsymbol{\theta}^{\hat{\tau}}, \boldsymbol{v}\rangle=\langle \boldsymbol{\theta}^{\hat{\tau}}, v^{\nu}\bold{e}_\nu\rangle = \sqrt{-g_{tt}}\,\dfrac{{\rm d}t}{{\rm d}\tau^\prime} = \sqrt{1+ \Pi^2/a(t)^2} = \gamma,
\end{equation}
where we have exploited the well-known relation $\langle \bold{d}x^\mu, \bold{e}_\nu \rangle = \delta^\mu_\nu$ and Eqs. (\ref{eq: dt/dtau_prime}),  (\ref{eq:theta_t}) and (\ref{eq:gamma_factor}).  Similarly, the photon energy $E_{\rm ph,local}$ measured by the Eulerian observer in his/her proper reference frame is given by
\begin{equation} \label{eq:E_p,local}
E_{\rm ph,local} = -\boldsymbol{k}\cdot \boldsymbol{n}=- k_{\hat{\tau}}=k^{\hat{\tau}}=  \sqrt{-g_{tt}}\,\dfrac{{\rm d}t}{{\rm d}s}=\dfrac{\vert \Pi_{\rm ph} \vert }{a(t)},
\end{equation}
$s$ being the affine parameter along the photon's null geodesics and $\Pi_{\rm ph}\equiv a(t)^2 {\rm d}x /{\rm d}s$ the conserved momentum. Therefore, the total energy of the traveller and the photon reads as, respectively,
\bsubeqs \label{eq:energy_total_traveller&photon}
\begin{equation} \label{eq:E_traveller}
E=-v_t=\sqrt{-g_{tt}}\, E_{\rm local}= \sqrt{\dfrac{t^2}{t^2+b^2}} \left(\sqrt{1 + \Pi^2/a(t)^2}\right),
\end{equation}
\begin{equation} \label{eq:E_photon}
E_{\rm ph}=-k_t=\sqrt{-g_{tt}}\, E_{\rm ph,local}= \sqrt{\dfrac{t^2}{t^2+b^2}} \left(\dfrac{\vert \Pi_{\rm ph} \vert}{a(t)}\right).
\end{equation}
\esubeqs
Equations (\ref{eq:E_local})--(\ref{eq:energy_total_traveller&photon}) show that even if we reject our proposal of defining the total energy according to Eq. (\ref{eq:Energy_tot}), we can at least say that the total energy $E$ of the traveller and the total energy $E_{\rm ph}$ of the photon make sense  at great distances from the defect (i.e., if $\vert t \vert \gg b$), where they reduce to $E_{\rm local}$ and $E_{\rm ph,local}$, respectively. This is due to the fact that for large values of $t$ the lapse function (\ref{lapse_function}) is such that $N \to 1$, meaning that  there is no difference between the Eulerian velocity $\boldsymbol{n}$  and  the normal evolution (four-)vector $\boldsymbol{m}$  if $\vert t \vert \gg b$ (cf. Eqs.  (\ref{eq:Eulerian_velocity_with_lapse})--(\ref{eq:modified_normal_evolution_vector})).  Therefore, we can write
\bsubeqs \label{eq:energies at great t}
\begin{equation}
E = -\boldsymbol{v} \cdot \boldsymbol{m} \xrightarrow[\vert t \vert \gg b]{} -\boldsymbol{v} \cdot \boldsymbol{n}= E_{\rm local},
\end{equation}
\begin{equation}
E_{\rm ph}=-\boldsymbol{k} \cdot \boldsymbol{m} \xrightarrow[\vert t \vert \gg b]{} -\boldsymbol{k} \cdot \boldsymbol{n}= E_{\rm ph,local}.
\end{equation}
\esubeqs

The geodesic equation provides us with the equations governing the dynamical evolution of $E$ and $E_{\rm ph}$. After an easy calculation, we obtain 
\bsubeqs \label{eq:equations_total_energies}
\begin{equation} \label{eq:equation for E_traveller}
\dot{E} +\dfrac{\dot{a}}{a} E -\dfrac{\dot{a}}{a} \dfrac{1}{E}\left( \dfrac{t^2}{b^2+t^2}\right) = \Gamma^{t}_{\phantom{t}tt} E,
\end{equation}
\begin{equation}\label{eq:equation for E_photon}
\dot{E}_{\rm ph} +\dfrac{\dot{a}}{a} E_{\rm ph} = \Gamma^{t}_{\phantom{t}tt} E_{\rm ph}.
\end{equation}
\esubeqs
A comparison with standard RW cosmology allows us to derive the equations ruling the dynamical behaviour of $E_{\rm local}$ and $E_{\rm ph,local}$, i.e., 
\bsubeqs \label{eq:equations_intrinsic_energies}
\begin{equation} \label{eq:equation for E_local}
\dot{E}_{\rm local} +\dfrac{\dot{a}}{a} E_{\rm local} -\dfrac{\dot{a}}{a} \dfrac{1}{E_{\rm local}} = 0,
\end{equation}
\begin{equation} \label{eq:equation for E_p,local}
\dot{E}_{\rm ph,local} +\dfrac{\dot{a}}{a} E_{\rm ph,local} =0.
\end{equation}
\esubeqs
The main difference between Eqs. (\ref{eq:equations_total_energies}) and (\ref{eq:equations_intrinsic_energies})  stems  from the  presence on the right hand side of  the former  of the connection coefficient $\Gamma^{t}_{\phantom{t}tt}$. As  anticipated before, we will interpret this term as due to  the (anti-)gravitational action generated by the defect. Furthermore, we  note in Eqs. (\ref{eq:equation for E_traveller}) and (\ref{eq:equation for E_local}) a contribution proportional to the inverse of the energy originating from the nonvanishing magnitude of traveller's four-velocity $\boldsymbol{v}$ via the condition $\boldsymbol{v}\cdot \boldsymbol{v}=-1$.

\begin{figure*}[th!]
\centering
\includegraphics[scale=0.74]{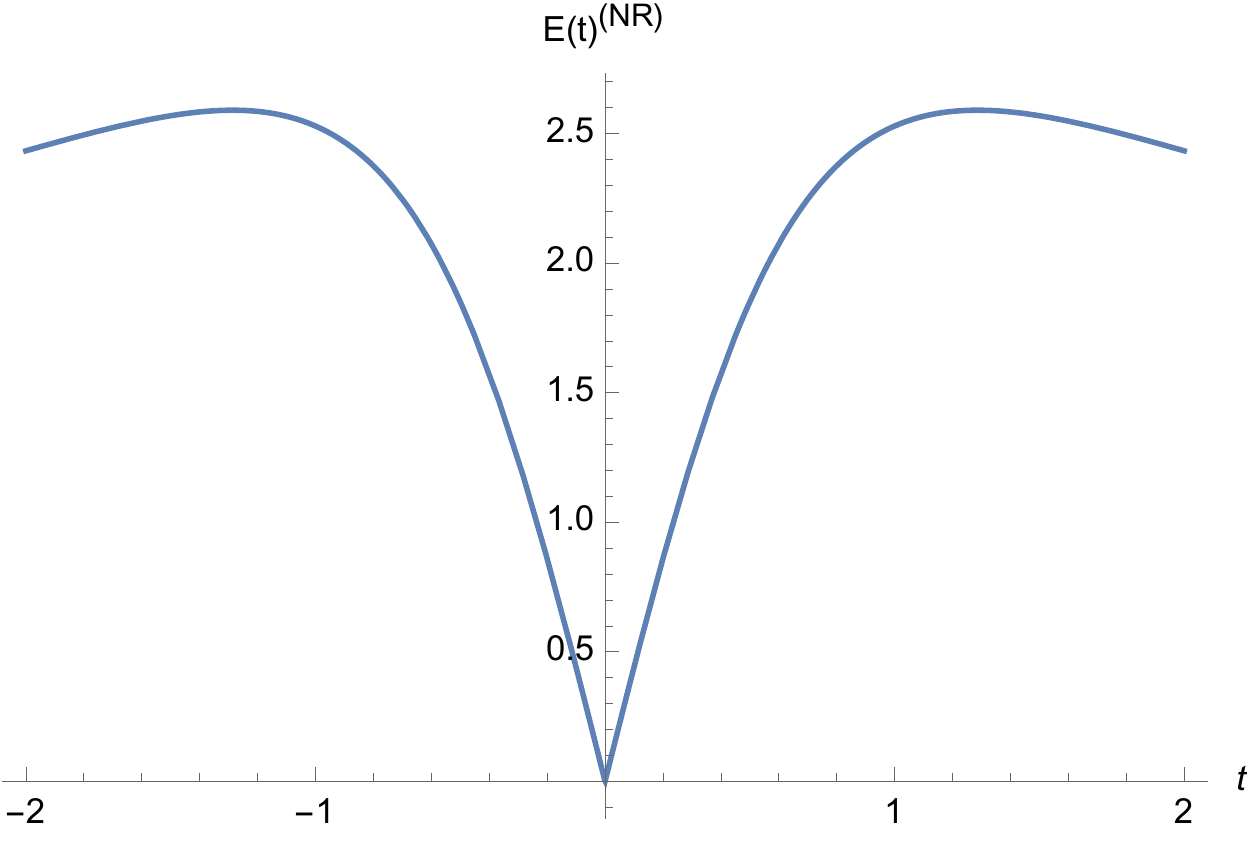}
\caption{The traveller energy (\ref{eq:E_traveller}) for the nonrelativistic-matter solution  (\ref{eq:scalefactor}) and with $\Pi=1$, $b=1$, $t_0=4 \sqrt{5}$. The suffix \qm{NR} stands for \qm{nonrelativistic}  cosmological matter content.}
\label{fig:E_traveller_non_relat}
\end{figure*}

\begin{figure*}[th!]
\centering
\includegraphics[scale=0.74]{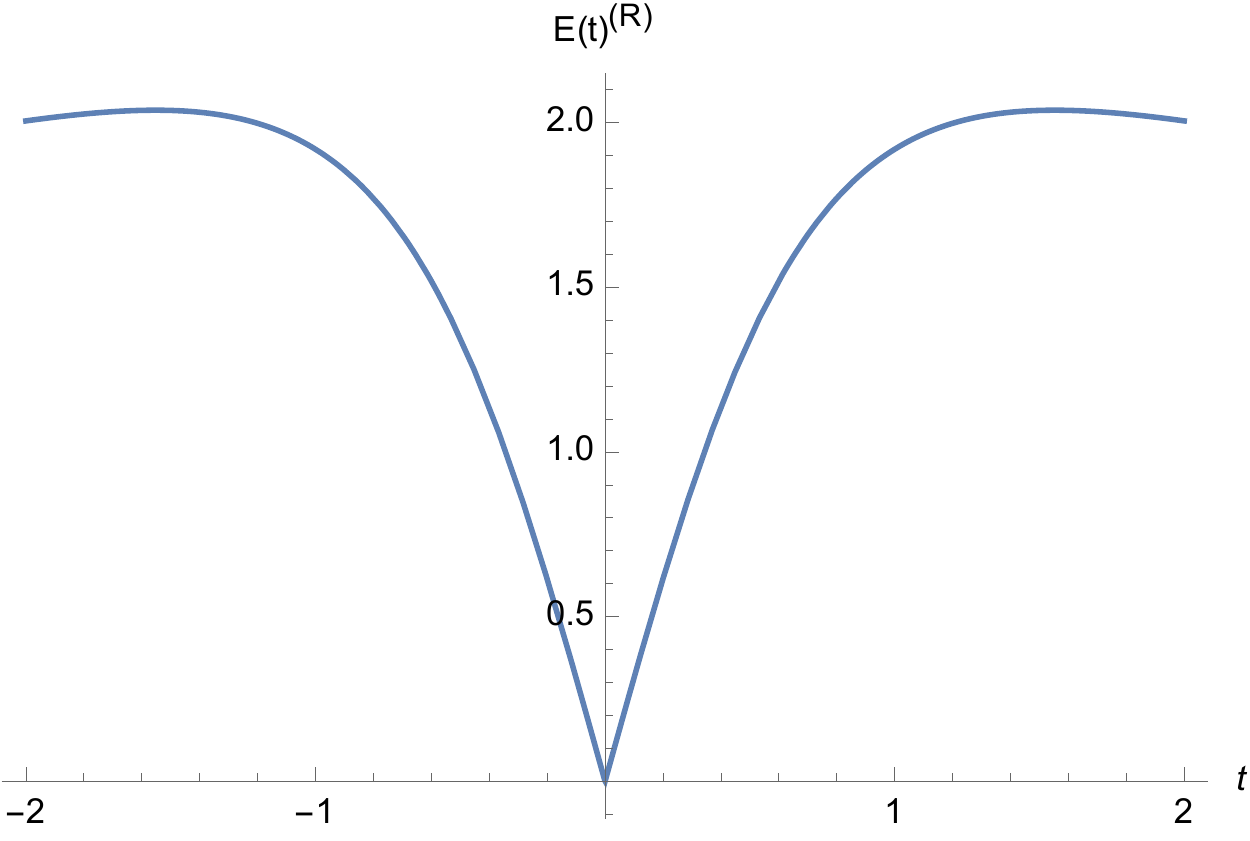}
\caption{The traveller energy (\ref{eq:E_traveller}) for the relativistic-matter solution  (\ref{eq:scalefactor}) and with $\Pi=1$, $b=1$, $t_0=4 \sqrt{5}$. The suffix \qm{R} stands for \qm{relativistic}  cosmological matter content.}
\label{fig:E_traveller_relat}
\end{figure*}

\begin{figure*}[th!]
\centering
\includegraphics[scale=0.74]{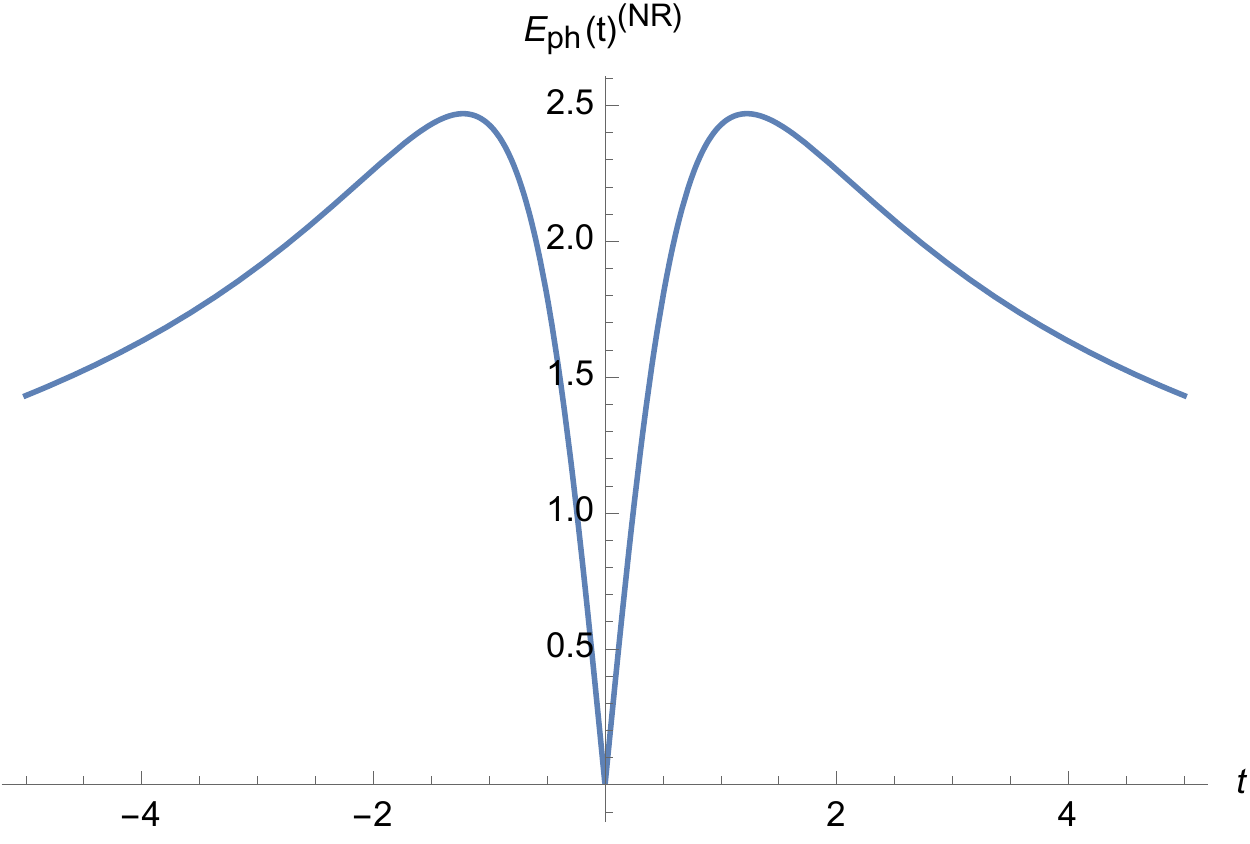}
\caption{The photon energy (\ref{eq:E_photon}) for the nonrelativistic-matter solution  (\ref{eq:scalefactor}) and with $\Pi_{\rm ph}=1$, $b=1$, $t_0=4 \sqrt{5}$. The suffix \qm{NR} stands for \qm{nonrelativistic}  cosmological matter content.}
\label{fig:E_photon_non_relat}
\end{figure*}

\begin{figure*}[th!]
\centering
\includegraphics[scale=0.74]{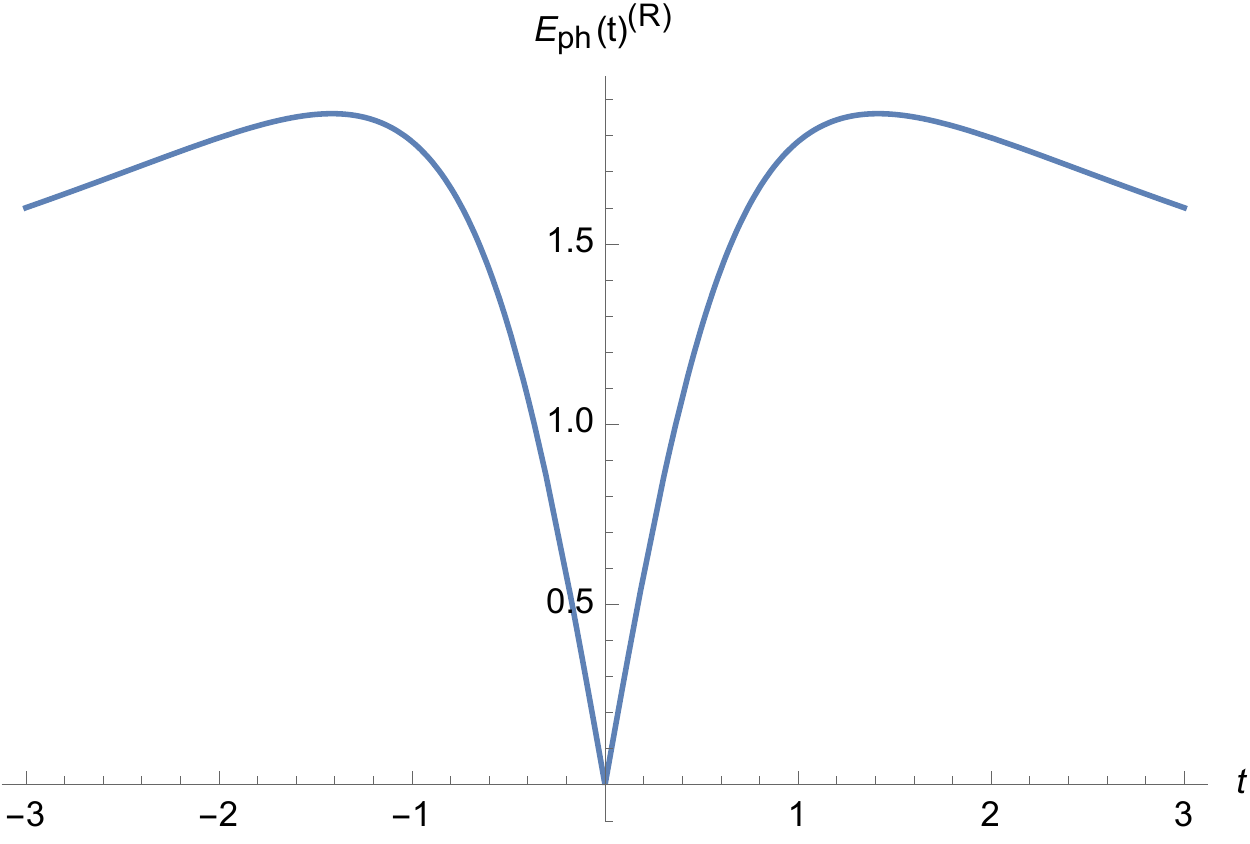}
\caption{The photon energy (\ref{eq:E_photon}) for the relativistic-matter solution  (\ref{eq:scalefactor}) and with $\Pi_{\rm ph}=1$, $b=1$, $t_0=4 \sqrt{5}$. The suffix \qm{R} stands for \qm{relativistic}  cosmological matter content.}
\label{fig:E_photon_relat}
\end{figure*}

The plots of the traveller energy $E$ and the photon energy $E_{\rm ph}$ are shown in Figs. \ref{fig:E_traveller_non_relat}--\ref{fig:E_photon_relat}.
From Figs.  \ref{fig:E_traveller_non_relat} and \ref{fig:E_traveller_relat}, it is clear that: in the negative-$t$ branch,  the total  energy of the traveller (\ref{eq:E_traveller}) starts decreasing after a time interval during which  it has increased; for $t>0$ we have the mirrored behaviour with respect to $t<0$; for  $t=0$ the energy vanishes and $\lim\limits_{t\to \pm \infty} E=1$ (corresponding to the traveller's rest mass). Moreover, apart from $t=0$, the energy (\ref{eq:E_traveller}) is always positive. The traveller energy has thus a usual (i.e., standard) behaviour only away from the defect, where it increases (resp. decreases) while the universe contracts (resp. expands); on the contrary, near the defect the energy diminishes (resp. grows) as the universe undergoes a contracting (resp. expanding) era.   

Something similar happens for the photon energy (\ref{eq:E_photon}), as Figs. \ref{fig:E_photon_non_relat} and \ref{fig:E_photon_relat}  witness. In this case, $E_{\rm ph}$ is always positive except for $t=0$ and for large values of $\vert t\vert $, where it vanishes. In addition, the simple form  assumed by Eq. (\ref{eq:E_photon}) permits to evaluate readily the sign of the time derivative $\dot{E}_{\rm ph}$. Indeed we find that (see Eq. (\ref{eq:t_star}))
\bsubeqs \label{eq:sign_of_dot_E_p}
\begin{equation}
\dot{E}_{\rm ph} >0 \Leftrightarrow  t<-\tilde{t}  \;\;  \cup \; \; 0<t<\tilde{t},
\end{equation}
\begin{equation}
\tilde{t} \equiv  \left \{
\begin{array}{rl} 
& t^\star/\sqrt{2}, \quad \quad {\rm nonrelativistic\; matter},\\
& t^\star, \quad \quad \quad \quad {\rm relativistic\; matter}.
\end{array} \right.
\end{equation}
\esubeqs
Following Eq. (\ref{eq:sign_of_dot_E_p}), we see that the photon energy  (\ref{eq:E_photon}) increases when $t<-\tilde{t}$ or $0<t<\tilde{t}$ and decreases if $-\tilde{t}<t<0$ or $t>\tilde{t}$. Thus, as for the traveller energy, $E_{\rm ph}$ has the usual behaviour only for $\vert t \vert \gg b$.

From the above analysis it is thus clear that the unusual character of the total energies $E$ and $E_{\rm ph}$  is due to the  (anti-)gravitational action exerted by the defect through the term $\Gamma^{t}_{\phantom{t}tt}$, as Eq. (\ref{eq:equations_total_energies}) shows. Furthermore, the same nonstandard behaviour affects also the Eulerian observer energy $\mathcal{E}$ being, as it is clear  from Eq. (\ref{eq:Eulerian_energy}), monotonically decreasing for $t < 0$ and monotonically increasing for $t > 0$ (indeed the shape of the energy function $\mathcal{E}$ resembles the traveller energy displayed in Figs. \ref{fig:E_traveller_non_relat} and \ref{fig:E_traveller_relat}). 

We also note that we can interpret as a discontinuity in the (instantaneous) power the discontinuity of first kind  that 
the total energy of the Eulerian observer, the traveller and the photon, Eqs. (\ref{eq:Eulerian_energy}) and (\ref{eq:energy_total_traveller&photon}), shows at at $t=0$ (see Figs. \ref{fig:E_traveller_non_relat}--\ref{fig:E_photon_relat}).

\begin{figure*}[th!]
\centering
\includegraphics[scale=0.74]{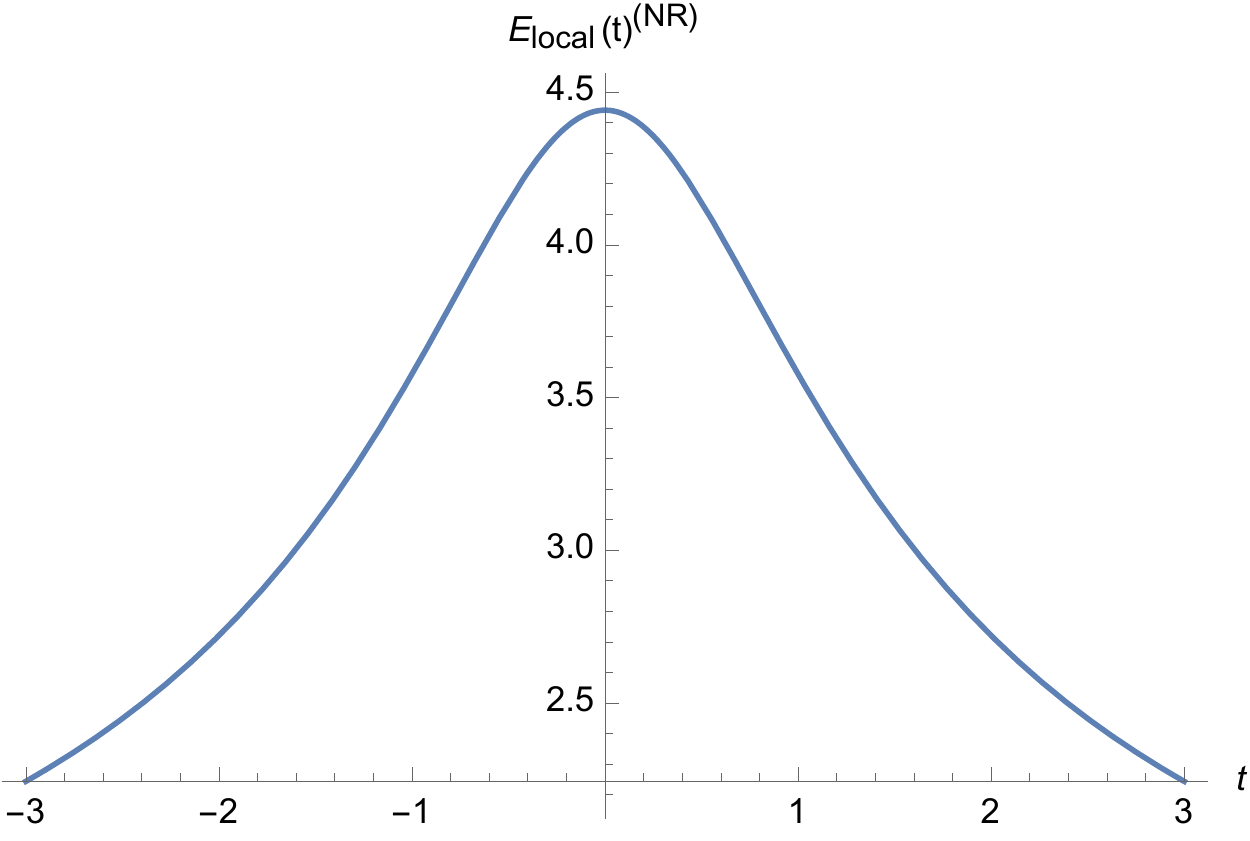}
\caption{The traveller energy (\ref{eq:E_local}) for the nonrelativistic-matter solution (\ref{eq:scalefactor}) and with $\Pi=1$, $b=1$, $t_0=4 \sqrt{5}$. It is clear that the maximum is reached at $t=0$. The suffix \qm{NR} stands for \qm{nonrelativistic} cosmological matter content.}
\label{fig:E_local_non_relat}
\end{figure*}

\begin{figure*}[th!]
\centering
\includegraphics[scale=0.74]{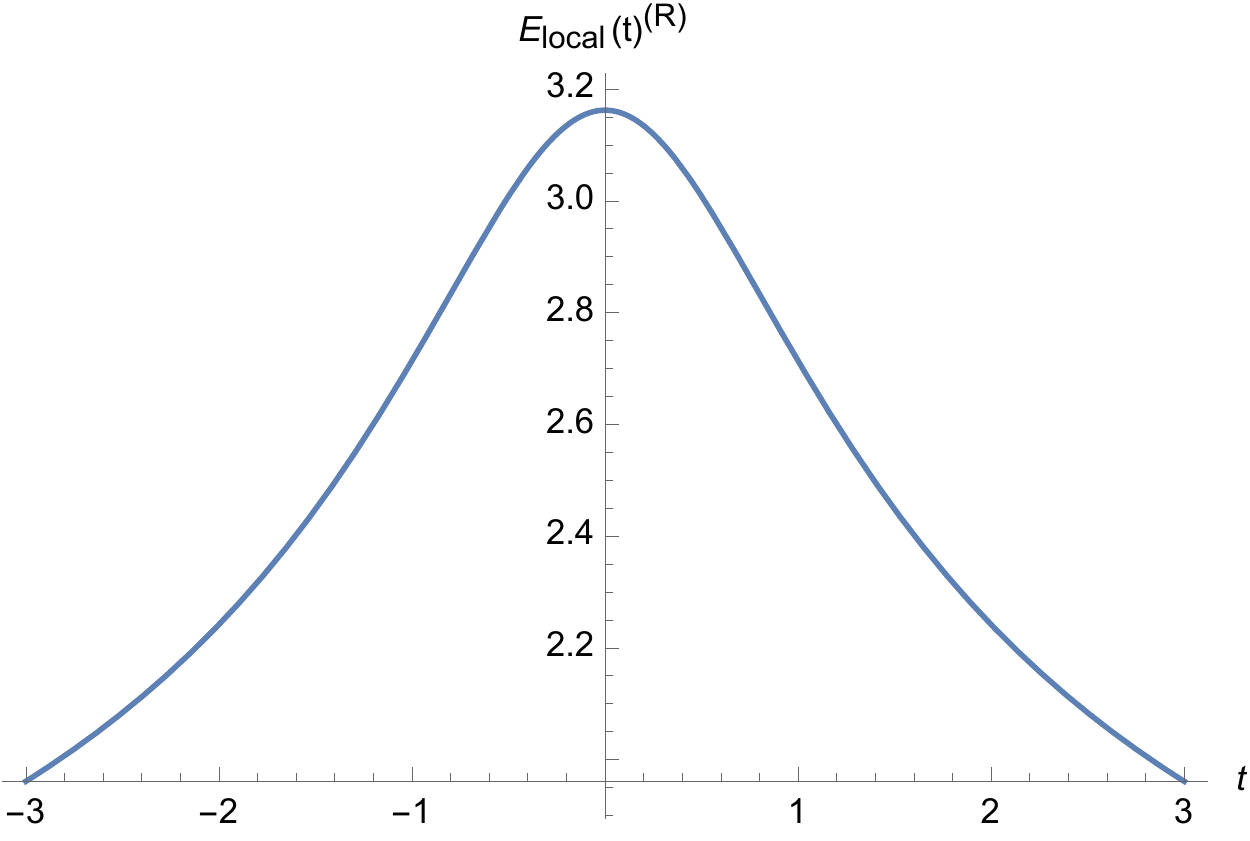}
\caption{The traveller energy (\ref{eq:E_local}) for the relativistic-matter solution  (\ref{eq:scalefactor}) and with $\Pi=1$, $b=1$, $t_0=4 \sqrt{5}$. It is clear that the maximum is attained when $t=0$. The suffix \qm{R} stands for \qm{relativistic}  cosmological matter content.}
\label{fig:E_local_relat}
\end{figure*}

At this stage, we can consider the traveller energy $E_{\rm local}$ and the photon energy $E_{\rm ph,local}$ as measured by the Eulerian observer in his/her proper reference frame $(\bold{e}_{\hat{\tau}},\bold{e}_{\hat{x}},\bold{e}_{\hat{y}},\bold{e}_{\hat{z}})$ (see Eqs. (\ref{eq:E_local}) and (\ref{eq:E_p,local})). Their plots are shown in Figs. \ref{fig:E_local_non_relat} and \ref{fig:E_local_relat} and Figs. \ref{fig:E_p,local_non_relat} and \ref{fig:E_p,local_relat}, respectively. It is thus clear 
that the energies (\ref{eq:E_local}) and (\ref{eq:E_p,local}) have always the usual character since, as we will show  below, no contribution from $\Gamma^{t}_{\phantom{t}tt}$ can be measured by the Eulerian observer in his/her proper reference frame (however, a first clear evidence is given by Eq. (\ref{eq:equations_intrinsic_energies})).
\begin{figure*}[th!]
\centering
\includegraphics[scale=0.74]{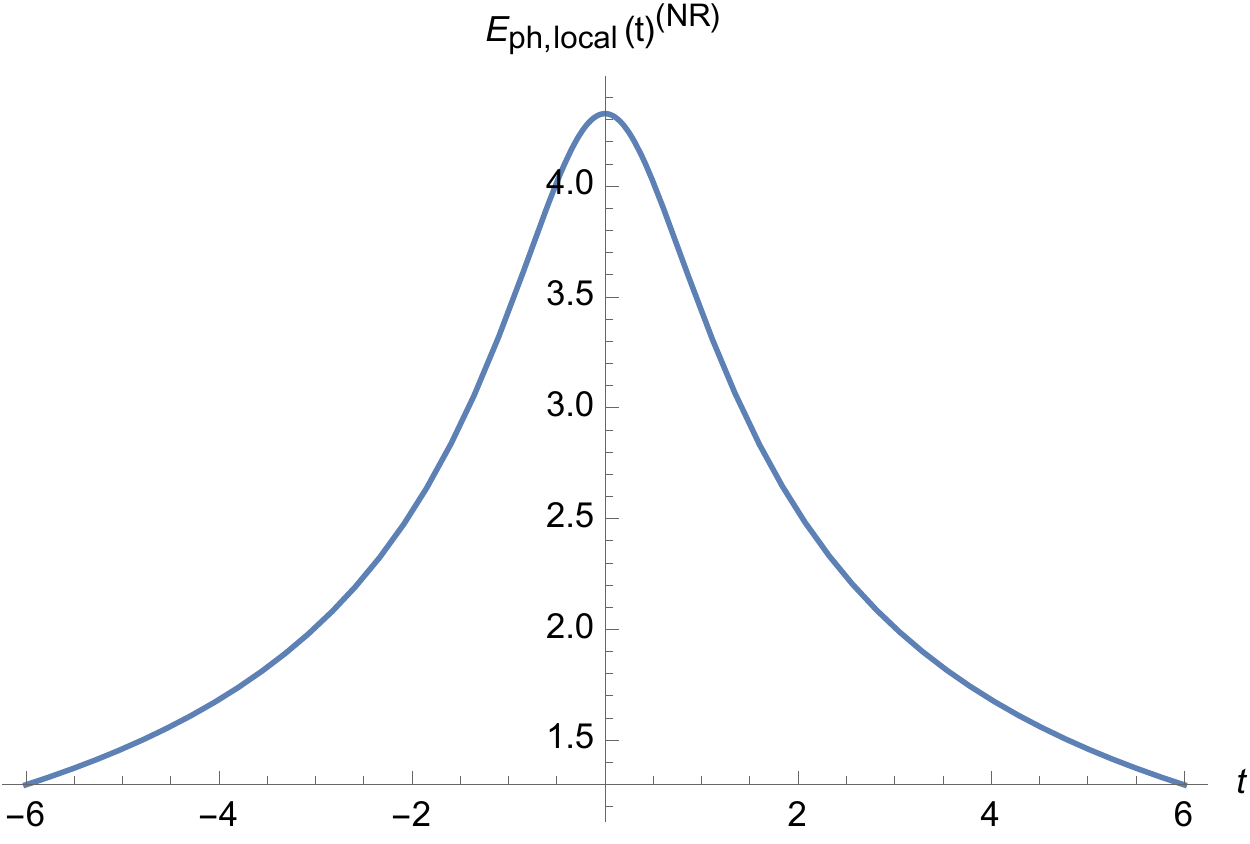}
\caption{The photon energy (\ref{eq:E_p,local}) for the nonrelativistic-matter solution  (\ref{eq:scalefactor}) and with $\Pi_{\rm ph}=1$, $b=1$, $t_0=4 \sqrt{5}$. It is clear that the maximum is reached at $t=0$. The suffix \qm{NR} stands for \qm{nonrelativistic}  cosmological matter content.}
\label{fig:E_p,local_non_relat}
\end{figure*}
\begin{figure*}[th!]
\centering
\includegraphics[scale=0.74]{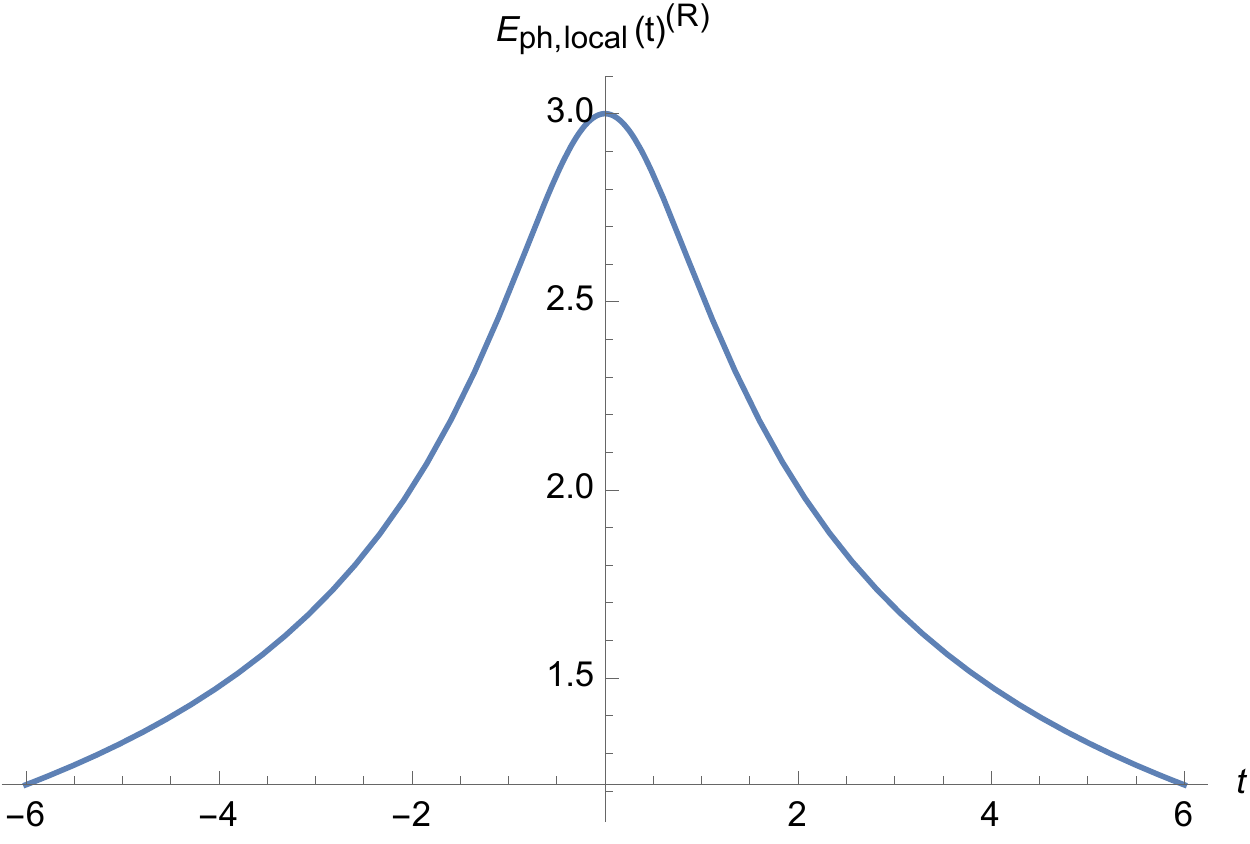}
\caption{The photon energy (\ref{eq:E_p,local}) for the relativistic-matter solution  (\ref{eq:scalefactor}) and with $\Pi_{\rm ph}=1$, $b=1$, $t_0=4 \sqrt{5}$. It is clear that the maximum is reached at $t=0$. The suffix \qm{R} stands for \qm{relativistic}  cosmological matter content.}
\label{fig:E_p,local_relat}
\end{figure*}
By employing hypotheses (\ref{eq:hypothesis}), we have that
\begin{equation} \label{eq:lim_E_local}
\lim_{t \rightarrow 0} E_{\rm local} =\sqrt{1+ \Pi^2/a(0)^2} \approx \left \{
\begin{array}{rl} 
& \vert \Pi \vert (t_0^2)^{1/3}/(b^2)^{1/3},  \quad \quad \quad \quad \quad {\rm nonrelativistic\; matter},\\
& \vert \Pi \vert (t_0^2)^{1/4}/(b^2)^{1/4}, \quad \quad \quad \quad \quad {\rm relativistic\; matter},
\end{array} \right.
\end{equation}
and hence it is clear that, at $t=0$, $E_{\rm local}$ is proportional to a fractional power of $1/b^2$ (see Figs. \ref{fig:E_local_non_relat} and \ref{fig:E_local_relat}). This signifies an enormous amount of energy, if we suppose that $b$ is proportional to the Planck length $\ell_{\rm P}$. The same conclusions  as those expressed by Eq. (\ref{eq:lim_E_local}) hold also for the photon energy $E_{\rm ph,local}$ if we  enforce (\ref{eq:hypothesis_1}).

We have already seen (cf. Eq. (\ref{eq:energies at great t})) that for large values of $t$ the total energies $E$ and $E_{\rm ph}$ reduce to the corresponding local intrinsic quantities $E_{\rm local}$ and $E_{\rm p, local}$, respectively. At this stage, we can provide a further explanation of this point. If we first  consider the case of photon energy, we can see that if $\vert t \vert \gg b$ then, from Eq. (\ref{eq:Gamma_sec_2}), $\Gamma^{t}_{\phantom{t}tt}$ tends to zero and hence Eq. (\ref{eq:equation for E_photon}) reduces to (\ref{eq:equation for E_p,local}), whose solution is represented by (\ref{eq:E_p,local}). The same holds also for the traveller energy,  where in addition the  term occurring in Eq. (\ref{eq:equation for E_traveller}), i.e., 
$$
\dfrac{t^2}{t^2+b^2},
$$
tends to one for large times and hence we can recover (\ref{eq:equation for E_local}), which in turn yields the expression (\ref{eq:E_local}).

As pointed out before, we can  interpret $\Gamma^{t}_{\phantom{t}tt}$ as the term embodying the (anti-)gravitational action exerted by the defect. At this stage, we can account for this assumption. First of all, we have already explained that $\Gamma^{t}_{\phantom{t}tt}$ fulfils an important role in the equations governing the dynamical evolution of the  total energy  of the traveller and the photon (cf. Eq. (\ref{eq:equations_total_energies})). Moreover, the fact that $\Gamma^{t}_{\phantom{t}tt}$ gets closer to zero if $\vert t \vert \gg b$ reflects   that the action of the defect phases out as a particle  travels away from it. Furthermore, $\Gamma^{t}_{\phantom{t}tt}$ diverges if $t \to 0$: the closer a particle  gets to the defect, the stronger is the \qm{force} it experiences. Moreover,  it is possible to write $\Gamma^{t}_{\phantom{t}tt}$  as (cf. Eq. (\ref{eq:Gamma_sec_2}))
\begin{equation} \label{eq:expression_of_Gamma_ttt}
\Gamma^{t}_{\phantom{t}tt} =-\dfrac{1}{t} \dfrac{\rho_{\rm defect}}{\rho},
\end{equation}
where (see Eq. (2.4c) in Ref. \cite{KWIII})
\begin{equation}
\rho_{\rm defect} \equiv -\dfrac{b^2 \rho}{b^2+t^2}.
\end{equation}
This means that the (anti-)gravitational \qm{force} produced by the defect depends on the ratio between the effective energy density $\rho_{\rm defect}$ of the defect and the  energy density $\rho$ of matter (or equivalently the ratio between their  masses) as well as the inverse time separation from the defect.

The above  analysis indicates also that $E$ and $E_{\rm ph}$ can be regarded as \emph{total} energies since they take into account also the contribution due to $\Gamma^{t}_{\phantom{t}tt}$, unlike the \emph{intrinsic} quantities $E_{\rm local}$ and $E_{\rm ph,local}$. Indeed, once  $\Gamma^{t}_{\phantom{t}tt}$  vanishes, the energy will not contain the contributions coming from the gravitational action of the defect but it will include only those terms connected  to the kinetic energy and the inertia. As a consequence,  when $\vert t \vert \gg b$ the total energies  reduce to the corresponding  intrinsic expressions, as we have just demonstrated.

We have already seen that the proper reference frame of the Eulerian observer is a freely falling frame whose coordinates amount to be Fermi normal coordinates. In such a frame,  the (Fermi normal) coordinates $x^{\hat{\mu}}$ of a generic point $\mathcal{P}$ located near the Eulerian observer's worldline are given  by
\begin{equation}
x^{\hat{\mu}}\left(\mathcal{P}\right)=\left(x^{\hat{\tau}},x^{\hat{j}}\right)= \left(\tau,s\,r^{\hat{j}}\right),
\end{equation}
where $\boldsymbol{r} = r^{\hat{j}}\bold{e}_{\hat{j}}$ is the tangent vector to the (spacelike) geodesic originating from the Eulerian observer's worldline at the specific (Eulerian observer's) proper time $\tau$ and $s$ the proper length along such geodesic. If we employ the notation $f(x^{\hat{\tau}}=\tau,x^{\hat{j}}=0) \equiv \left.f\right\vert_{\mathcal{G}}$ to indicate that a quantity is evaluated along the Eulerian observer's geodesic, we know from Ref. \cite{Manasse-Minser1963} that Christoffel symbols satisfy the following relations:
\bsubeqs
\begin{equation}
\left. \partial_{\hat{\tau}} \Gamma^{\hat{\alpha}}_{\phantom{\hat{\alpha}}\hat{\mu}\hat{\nu}} \right\vert_{\mathcal{G}} = 0,
\end{equation}
\begin{equation}
\left. \partial_{\hat{\nu}} \Gamma^{\hat{\alpha}}_{\phantom{\hat{\alpha}}\hat{\mu}\hat{\tau}} \right\vert_{\mathcal{G}} =\left. R^{\hat{\alpha}}_{\phantom{\hat{\alpha}}\hat{\mu}\hat{\nu}\hat{\tau}} \right\vert_{\mathcal{G}},
\end{equation}
\begin{equation}
\left. \partial_{\hat{k}} \Gamma^{\hat{\alpha}}_{\phantom{\hat{\alpha}}\hat{i}\hat{j}} \right\vert_{\mathcal{G}} = -\dfrac{1}{3} \left(\left. R^{\hat{\alpha}}_{\phantom{\hat{\alpha}}\hat{i}\hat{j}\hat{k}} \right\vert_{\mathcal{G}} +\left. R^{\hat{\alpha}}_{\phantom{\hat{\alpha}}\hat{j}\hat{i}\hat{k}} \right\vert_{\mathcal{G}} \right),
\end{equation}
\esubeqs
from which we derive the following expansion for the Christoffel symbols:
\begin{equation} \label{eq:Christoffel_fermi_normal_coordinates}
\Gamma^{\hat{\alpha}}_{\phantom{\hat{\alpha}}\hat{\mu}\hat{\nu}} \left(\tau,x^{\hat{j}}\right)= \left. \partial_{\hat{k}} \Gamma^{\hat{\alpha}}_{\phantom{\hat{\alpha}}\hat{\mu}\hat{\nu}} \right\vert_{\mathcal{G}} x^{\hat{k}} + {\rm O} \left( \vert x^{\hat{k}} \vert^2\right).
\end{equation}
From the above equation, we obtain the following relation, valid for our modified RW model: 
\begin{equation} \label{eq:Gamma_000_cappuccio_expansion}
\Gamma^{\hat{\tau}}_{\phantom{\hat{\tau}}\hat{\tau}\hat{\tau}} \left(\tau,x^{\hat{j}}\right)=  {\rm O} \left( \vert x^{\hat{k}} \vert^2\right). 
\end{equation}
This means that the connection coefficient $\Gamma^{\hat{\tau}}_{\phantom{\hat{\tau}}\hat{\tau}\hat{\tau}}$ vanishes  not only along the geodesic $\mathcal{G}$ of the Eulerian observer but also  in the neighbourhood of $\mathcal{G}$ (within the precision of ${\rm O} ( \vert x^{\hat{k}} \vert)$). Accordingly, the energies $E_{\rm local}$ and $E_{\rm ph,local}$ measured by the Eulerian observer in his/her proper reference frame will not include the effects coming from $\Gamma^{\hat{\tau}}_{\phantom{\hat{\tau}}\hat{\tau}\hat{\tau}}$. This offers another explanation of the fact that,  once the underlying computations are performed in the coordinate system $(t,x,y,z)$, the dynamical evolution of $E_{\rm local}$ and $E_{\rm ph,local}$ are ruled by Eq.  (\ref{eq:equations_intrinsic_energies}), where no contribution from  $\Gamma^{t}_{\phantom{t}tt}$ appears. In other words, the Eulerian observer will not take into account the gravitational action of the defect (represented, as we said before, by $\Gamma^{t}_{\phantom{t}tt}$) when he/she measures the energy of the photon or the massive particle.  

We can also explain this point with a more direct approach. We know that, in Fermi normal coordinates, the geodesic equation can be written as
\begin{equation} \label{eq:geodesic_equation_fermi_normal_coordinates}
\dfrac{{\rm d^2}x^{\hat{\mu}}}{{\rm d}\lambda^2}+\Gamma^{\hat{\mu}}_{\phantom{\hat{\mu}}\hat{\nu}\hat{\alpha}} \left(\tau,x^{\hat{j}}\right) \dfrac{{\rm d}x^{\hat{\nu}}}{{\rm d}\lambda}\dfrac{{\rm d}x^{\hat{\alpha}}}{{\rm d}\lambda}=0,
\end{equation}
where $\lambda$ is the affine parameter along the geodesic (which in the case of timelike geodesic  can be identified with the particle's proper time).  Therefore, the energy $p^{\hat{\tau}}$ of a massive particle or a photon having four-momentum $p^{\hat{\alpha}}= {\rm d}x^{\hat{\alpha}}/{\rm d}\lambda$ as measured by the Eulerian observer will obey the equation (we  simply write $\Gamma^{\hat{\tau}}_{\phantom{\hat{\tau}}\hat{\nu}\hat{\alpha}}(\tau,x^{\hat{j}})   \equiv \Gamma^{\hat{\tau}}_{\phantom{\hat{\tau}}\hat{\nu}\hat{\alpha}}$ in order to ease the notation)
\begin{equation} \label{eq:0_compon_geodesic_equation_fermi_normal_coordinates}
\dfrac{{\rm d}p^{\hat{\tau}}}{{\rm d}\lambda}+\Gamma^{\hat{\tau}}_{\phantom{\hat{\tau}}\hat{\alpha}\hat{\beta}}  p^{\hat{\alpha}} p^{\hat{\beta}}= 0.
\end{equation}
Bearing in mind  Eqs. (\ref{eq:cappuccio_first})--(\ref{eq: e_alpha_cappuccio}) and (\ref{eq:gamma_hat_indeces}), we find that the coefficients $\Gamma^{\hat{\tau}}_{\phantom{\hat{\tau}}\hat{\alpha}\hat{\beta}}$ occurring in Eq. (\ref{eq:0_compon_geodesic_equation_fermi_normal_coordinates}) are given by
\bsubeqs \label{eq:Gamma^0_mu-nu cappuccio}
\begin{equation} \label{eq:Gamma_000_cappuccio}
\Gamma^{\hat{\tau}}_{\phantom{\hat{\tau}}\hat{\tau}\hat{\tau}}= \partial_t e_{\hat{\tau}}^{\phantom{\hat{\tau}}t}+e_{\hat{\tau}}^{\phantom{\hat{\tau}}t} \Gamma^{t}_{\phantom{t}tt} =0,
\end{equation}
\begin{equation}
\Gamma^{\hat{\tau}}_{\phantom{\hat{\tau}}\hat{\tau}\hat{i}} = e_{\hat{i}}^{\phantom{\hat{i}}j} \Gamma^{t}_{\phantom{t}tj}=0,
\end{equation}
\begin{equation}
\Gamma^{\hat{\tau}}_{\phantom{\hat{\tau}}\hat{i}\hat{\tau}}=  e^{\hat{\tau}}_{\phantom{\hat{\tau}}t} e_{\hat{i}}^{\phantom{\hat{i}}i} \left(\partial_i e_{\hat{\tau}}^{\phantom{\hat{\tau}}t}+e_{\hat{\tau}}^{\phantom{\hat{\tau}}t} \Gamma^{t}_{\phantom{t}ti}\right)=0,
\end{equation}
\begin{equation}
\Gamma^{\hat{\tau}}_{\phantom{\hat{\tau}}\hat{i}\hat{j}}=  e^{\hat{\tau}}_{\phantom{\hat{\tau}}t} e_{\hat{i}}^{\phantom{\hat{i}}i} e_{\hat{j}}^{\phantom{\hat{j}}j} \Gamma^{t}_{\phantom{t}ij} = \sqrt{\dfrac{b^2+t^2}{t^2}}\dfrac{\dot{a}(t)}{a(t)} \delta_{\hat{i}\hat{j}}.
\end{equation}
\esubeqs
By means of  Eq. (\ref{eq:Gamma^0_mu-nu cappuccio}), we can show that Eq. (\ref{eq:0_compon_geodesic_equation_fermi_normal_coordinates}) reduces to Eq. (\ref{eq:equations_intrinsic_energies}) once all calculations are performed in $(t,x,y,z)$ coordinates. However, the most important point of this computation is  that Eq. (\ref{eq:Gamma^0_mu-nu cappuccio}) clearly shows that $\Gamma^{\hat{\tau}}_{\phantom{\hat{\tau}}\hat{\tau}\hat{\tau}}$ is the only connection coefficient occurring in Eq. (\ref{eq:0_compon_geodesic_equation_fermi_normal_coordinates}) which depends on $\Gamma^{t}_{\phantom{t}tt}$. Since  $\Gamma^{\hat{\tau}}_{\phantom{\hat{\tau}}\hat{\tau}\hat{\tau}}$ vanishes,  no  contribution from $\Gamma^{t}_{\phantom{t}tt}$  appears in (\ref{eq:equations_intrinsic_energies}). As a result, the Eulerian observer will not measure the effects related to  $\Gamma^{t}_{\phantom{t}tt}$, i.e., those contributions we are interpreting as due to the (anti-)gravitational action of the defect.

At this stage, let us stress another point. We  have seen in Sec. \ref{Sec:Compressive_forces} that  compressive forces felt by the Eulerian (human) observer  show no deviations from the expectations of standard cosmology. It is now clear that this result is due to the fact that the effects introduced by $\Gamma^{t}_{\phantom{t}tt} $ cannot be measured in the freely falling frame $(\bold{e}_{\hat{\tau}},\bold{e}_{\hat{x}},\bold{e}_{\hat{y}},\bold{e}_{\hat{z}})$, see Eq. (\ref{eq:Riemann_hat_indices_3}) and comments below. Bearing in mind the previous analysis regarding the local energies (\ref{eq:E_local}) and (\ref{eq:E_p,local}), this means that physical quantities measured by the Eulerian observer do not  display an unusual behaviour because in the proper reference frame $(\bold{e}_{\hat{\tau}},\bold{e}_{\hat{x}},\bold{e}_{\hat{y}},\bold{e}_{\hat{z}})$ no contribution from $\Gamma^{t}_{\phantom{t}tt} $ can appear.

In our analysis a final question must be answered. For the sake of clarity, in the following calculations we will keep 
the observers' rest mass explicit. Accordingly, let $\mu_{\rm E}$ denote the Eulerian observer's rest mass.  We know that comoving coordinates $(t,x,y,z)$ introduced in Eqs. (\ref{t_coord}) and (\ref{x^i_coord}) allow for the description of  the modified model of universe from the point of view of the Eulerian observer. Recalling that such observer is always at rest in these  coordinates,  we are thus led to wonder about the reasons for which our definition  (\ref{eq:Energy_tot}) of total energy leads to Eq. (\ref{eq:energy_eulerian}) instead of  an expression like $\mathcal{E} = \mu_{\rm E}$. 
The answer is that  (\ref{eq:Energy_tot}) defines a total energy and hence it takes into account also the defect's gravitational action. This means that the total energy of the Eulerian observer  (\ref{eq:energy_eulerian}) (or equivalently (\ref{eq:Eulerian_energy})) receives a contribution from the defect such that $\mathcal{E} \neq \mu_{\rm E}$. Indeed, we simply have 
\begin{equation} \label{eq:mathcal_E_with_mass}
\mathcal{E}= -\mu_{\rm E} \left(\boldsymbol{n}\cdot \boldsymbol{m}\right)=\mu_{\rm E} \sqrt{\dfrac{t^2}{t^2+b^2}}= \sqrt{\mu_{\rm E}^2 + \mu_{\rm E}^2  \dfrac{\rho_{\rm{defect}}}{\rho}}.
\end{equation}
In other words, $\mathcal{E}$ differs from $\mu_{\rm E}$ by a term involving the ratio $\rho_{\rm defect}/\rho$, i.e., the same term occurring in the expression of $\Gamma^{t}_{\phantom{t}tt}$ (see Eq. (\ref{eq:expression_of_Gamma_ttt})). Therefore,  the presence of the defect makes $\mathcal{E}$ differ from the Eulerian observer's rest mass $\mu_{\rm E}$.  Similarly, the  total energy of the traveller (\ref{eq:E_traveller}) can be written as
\begin{equation}
E= -\mu_{\rm T} \left(\boldsymbol{v}\cdot \boldsymbol{m}\right)= \sqrt{\mu_{\rm T}^2 + \mu_{\rm T}^2  \dfrac{\rho_{\rm{defect}}}{\rho}} \sqrt{1 + \dfrac{\Pi^2}{a(t)^2}},
\end{equation}
$\mu_{\rm T}$ being the traveller's rest mass. In this expression we can recognize both the defect's gravitational action, represented by the term $\mu_{\rm T}  \rho_{\rm defect}/\rho$ and the \qm{kinetic} term proportional to $\Pi^2/a(t)^2$. Finally, the photon energy (\ref{eq:E_photon}) reads as  
\begin{equation} \label{eq:E_p_with_effective_mass}
E_{\rm ph}= \sqrt{1+\dfrac{\rho_{\rm defect}}{\rho}} \,\dfrac{\vert \Pi_{\rm ph} \vert }{a(t)}.
\end{equation}
In this case, we can interpret the \qm{correction} term $\rho_{\rm defect}/\rho$ as a photon's effective mass induced by the defect.

Equations (\ref{eq:mathcal_E_with_mass})--(\ref{eq:E_p_with_effective_mass}) support, once again, our proposal of interpreting (\ref{eq:Energy_tot}) as  the \emph{total} energy of a massive particle/photon with four-momentum $\boldsymbol{p}$. On the other hand, (\ref{eq:E_measured}) denotes an \emph{intrinsic} energy which does not take into account the defect's gravitational force, represented by the term $\Gamma^{t}_{\phantom{t}tt} $. Indeed, in Eqs. (\ref{eq:mathcal_E_with_mass})--(\ref{eq:E_p_with_effective_mass})  the term $\rho_{\rm defect}/\rho$, originating from $\Gamma^{t}_{\phantom{t}tt}$, appears explicitly. On the contrary, in Eqs. (\ref{eq:E_local}) and (\ref{eq:E_p,local}) no contribution coming from $\Gamma^{t}_{\phantom{t}tt}$ can occur, as we have explained before (cf. Eq. (\ref{eq:equations_intrinsic_energies}) and comments following  Eqs. (\ref{eq:Gamma_000_cappuccio_expansion}) and  (\ref{eq:Gamma^0_mu-nu cappuccio})).

\section{Conclusions and open problems} \label{Sec:Conclusions}

The main purpose of this paper consists in seeking physical observables which can point out the presence of the spacetime defect characterizing the regularized RW geometry (\ref{modified_FRW}), where the big-bang singularity has been tamed by a nonzero length parameter $b$. Our description relies on the analysis of two physical quantities: the compressive forces acting on (human) observers and the energy of massive particles and photons crossing it. The first topic has been explored in Sec. \ref{Sec:Compressive_forces}. We have devised a reasonable criterion to single out the defect, which can be defined as the three-dimensional hypersurface where  the \emph{modulus} of  compressive forces attain their maximum value (see Fig. \ref{fig:compressive_forces}). Furthermore, we have  found that if we take the proposal $b \sim {\ell}_{\rm P} $ made  in Appendix B of Ref. \cite{KII}  seriously, then  the defect can be modelled as a gravitational obstacle with compressive forces  proportional to $1/b$ (see Eqs. (\ref{eq:tidal_forces_Eulerian_2}), (\ref{eq:tidal_traveller}) and (\ref{eq:tidal_traveller_approx})). A rough calculation reveals that a Eulerian human observer can withstand compressive forces generated at $t=0$ if $b \gtrsim 10^7$ m.
This result can lead to interesting implications due to the possibility of having a quantum-inspired  defect length scale. In particular, we have explained how our investigation seems to agree with the scenario drawn in Ref. \cite{Klinkhamer2021a} suggesting  that a new physics phase at $t=0$ could create a pair of separated universes.

In Sec. \ref{Sec:Energy}, we have provided a definition of a \emph{total} energy suitable for our model (see comments accompanying Eqs. (\ref{eq:Energy_tot}) and (\ref{eq:energies at great t})) which  differentiates it from the (local) \emph{intrinsic} energy measured by the Eulerian observer (see comments below Eqs. (\ref{eq:expression_of_Gamma_ttt}) and (\ref{eq:E_p_with_effective_mass})). We have seen that both the total energy  of the generic freely falling observer and of the photon, defined according to (\ref{eq:Energy_tot}) and given in  Eq. (\ref{eq:energy_total_traveller&photon}), exhibit an unusual character over a finite time interval around  $t=0$: they grow (resp. diminish) as the universe expands (resp. contracts); see Figs. \ref{fig:E_traveller_non_relat}--\ref{fig:E_photon_relat}. As an inspection of Eq. (\ref{eq:equations_total_energies}) reveals, this scenario is due to the effects introduced by the Christoffel symbol $\Gamma^{t}_{\phantom{t}tt}$ (cf. Eq. (\ref{eq:Gamma_sec_2})), which we propose to interpret  as the term embodying the (anti-)gravitational action exerted by the defect (see comments below Fig. \ref{fig:E_p,local_relat}). Furthermore, the same nonstandard behaviour affects also the  Eulerian observer energy (\ref{eq:Eulerian_energy}). On the other hand, the intrinsic energy of the generic  freely falling traveller and of the photon,  given in Eqs.  (\ref{eq:E_local}) and (\ref{eq:E_p,local}), respectively,  displays no unusual property (see Figs. \ref{fig:E_local_non_relat}--\ref{fig:E_p,local_relat} and comments therein). This absence of discrepancy  with respect to standard cosmology predictions stems from the fact that the Eulerian observer cannot measure contributions related to the Christoffel symbol $\Gamma^{t}_{\phantom{t}tt}$ in his/her proper reference frame (see comments  following  Eqs. (\ref{eq:Gamma_000_cappuccio_expansion}) and  (\ref{eq:Gamma^0_mu-nu cappuccio})). 

\begin{figure*}[th!]
\centering
\includegraphics[scale=0.35]{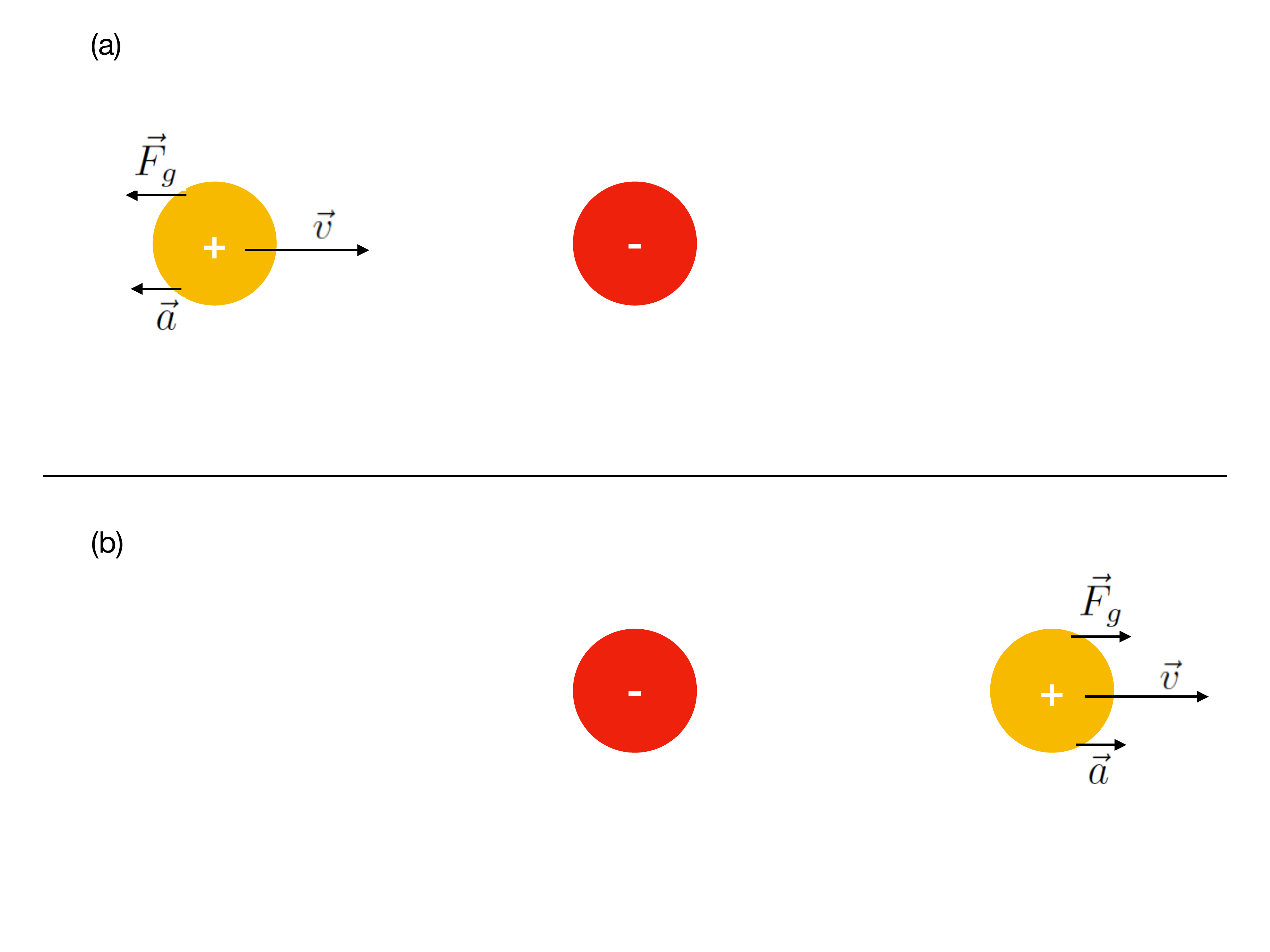}
\caption{Gravitational repulsion experienced by a positive (yellow)  mass due to the presence of a  body with negative (red) mass, which is supposed to be fixed. The interaction is described through Newtonian theory. (a) The positive mass goes toward the fixed body with  velocity $\vec{v}$; (b) the positive mass  moves away from the fixed body with  velocity $\vec{v}$. In both cases, the gravitational repulsive force $\vec{F}_g$ and the resulting acceleration $\vec{a}$ have the same direction and point away from the negative mass.}
\label{fig:antigravitational_interact}
\end{figure*}

A possible explanation of the nonstandard behaviour of the energy can come from the following hypothesis involving  gravitationally repulsive negative masses. A negative mass is an exotic matter which would violate one or more energy conditions. It can be implemented in general relativity theory \cite{Bondi1957}, where the equivalence principle implies that the inertial mass  equals the passive gravitational mass\footnote{Active and passive gravitational masses are identical due to the law of conservation of momentum.}.  Consider the situation depicted in Fig. \ref{fig:antigravitational_interact}, where, for simplicity, the gravitational positive-negative mass interaction  is explored by means of  Newtonian theory and  the negative mass is supposed to be fixed. The gravitational repulsive force $\vec{F}_g$ experienced by the moving body  and the resulting acceleration $\vec{a}$ point along the same direction. In panel (a), the positive mass   approaches with velocity $\vec{v}$ the negative mass. Since the work done by $\vec{F}_g$ is negative, the positive mass slows down. In  panel (b), the positive mass  departs  from the fixed body. In this case, it is clear that the  kinetic energy  of the moving body  increases. Furthermore, the mechanical energy is conserved in both situations. This classical example can give us some insight into the nature of the defect. First of all, recall that in a generic spacetime there will not be a well-defined notion of gravitational potential energy (although in special cases it  exists). In our  relativistic model,
if we conceive the defect as an object having  negative (active gravitational) mass, then it is possible to account for the behaviour of the energy as shown in Figs. \ref{fig:E_traveller_non_relat}--\ref{fig:E_photon_relat}. Indeed, over a small time interval around the bounce, where anti-gravitational phenomena  reveal an increasing importance,   the energy of both  massive particles and  photons decreases for negative values of the time variable $t$ and   grows when $t$ becomes positive. This reflects the behaviour of the kinetic energy in the classical example of Fig. \ref{fig:antigravitational_interact}. Moreover,  at great distances from the defect,  the energy displays its standard behaviour,  meaning that antigravitational effects are negligible (as witnessed by the fact that $\Gamma^{t}_{\phantom{t}tt}$ goes to zero if $\vert t \vert \gg b$, see Eq. (\ref{eq:Gamma_sec_2})). Furthermore, this scenario allows for the fact that the energy drops to zero at $t=0$: the defect drains the energy of particles until, at the bounce, it vanishes; after that, the defect gives to particles the required energy  to carry on with their motion. Therefore, we can conclude that the effective NEC violation featuring the defect and the related antigravity effects represent the source of the nonstandard behaviour of the total energy shown in Figs. \ref{fig:E_traveller_non_relat}--\ref{fig:E_photon_relat}. Incidentally, the possibility of modelling the defect as an object having negative  mass has been discussed also in Ref. \cite{Klinkhamer-Queiruga2018}. This hypothesis can open interesting prospectives. As an example, recently in Ref. \cite{Farnes2017} gravitationally repulsive negative  masses have been proposed  as natural candidates for the description of both dark matter and dark energy. Thus, we might wonder if also the defect can represent   such candidate. 

The analysis of Sec. \ref{Sec:Energy} has shown that the nonstandard properties  of massive particles and photons energy  are not confined to the single point $t=0$, but they concern a finite time interval around the defect's location which is of the order of the characteristic length scale $b$.  This pattern is in accordance with the aforementioned proposed interpretation of the  connection coefficient $\Gamma^{t}_{\phantom{t}tt}$ since, as we have explained in the comments above Eq. (\ref{eq:expression_of_Gamma_ttt}), this function  attains large values in a region around $t=0$ (except  for $t=0$, where it is  not defined) and approaches  zero as $\vert t \vert \gg b$.  This means that  the unusual energy's behaviour configures as  a sort of \qm{spreading effect} able to  encompass regions near the defect. This is not the first  example of a \qm{spreading effect} occurring in the modified RW geometry (\ref{modified_FRW}), the other one being represented by the effective NEC violation which, as reported in Refs. \cite{KII,KWIII},   can be extended to a finite interval around $t = 0$. Therefore,  these results lead quite naturally to one important (open) question: are there  other \qm{spreading phenomena} in nonsingular-bouncing-cosmology settings?

Another fascinating issue to be addressed, especially in light of the outcome of this paper, regards the origin of the defect length scale $b$. Indeed, if  its quantum origin were proven, then an intriguing task would consists in trying to reconcile this fact with the arguments spelled out in our paper. On the other hand, a broader investigation is performed   in Ref. \cite{K2020-IIB-3}, where  it is argued that  $b$ could be a remnant of a new (not necessarily quantum) physics phase  replacing Einstein gravity (see also Ref. \cite{Klinkhamer2021a}). It would be interesting to determine  if there exists a connection between the physical effects discussed in this paper and the new phase mentioned in Ref. \cite{K2020-IIB-3}.

Finally, the rich mathematical structure underlying regularized RW geometry (\ref{modified_FRW}) (likewise degenerate metrics in general, see Ref. \cite{Gunther}) still deserves  further investigation. Indeed,  this can represent, on the one hand, a way  to find other physical phenomena associated with the defect, and, on the other, we can expect to obtain equivalent explanations for the physical effects described in this paper.

\section*{Acknowledgements}

It is a pleasure to thank F. R. Klinkhamer for  extensive discussions.

\bibliography{references}

\begin{thebibliography}{30}%
\makeatletter
\providecommand \@ifxundefined [1]{%
 \@ifx{#1\undefined}
}%
\providecommand \@ifnum [1]{%
 \ifnum #1\expandafter \@firstoftwo
 \else \expandafter \@secondoftwo
 \fi
}%
\providecommand \@ifx [1]{%
 \ifx #1\expandafter \@firstoftwo
 \else \expandafter \@secondoftwo
 \fi
}%
\providecommand \natexlab [1]{#1}%
\providecommand \enquote  [1]{``#1''}%
\providecommand \bibnamefont  [1]{#1}%
\providecommand \bibfnamefont [1]{#1}%
\providecommand \citenamefont [1]{#1}%
\providecommand \href@noop [0]{\@secondoftwo}%
\providecommand \href [0]{\begingroup \@sanitize@url \@href}%
\providecommand \@href[1]{\@@startlink{#1}\@@href}%
\providecommand \@@href[1]{\endgroup#1\@@endlink}%
\providecommand \@sanitize@url [0]{\catcode `\\12\catcode `\$12\catcode
  `\&12\catcode `\#12\catcode `\^12\catcode `\_12\catcode `\%12\relax}%
\providecommand \@@startlink[1]{}%
\providecommand \@@endlink[0]{}%
\providecommand \url  [0]{\begingroup\@sanitize@url \@url }%
\providecommand \@url [1]{\endgroup\@href {#1}{\urlprefix }}%
\providecommand \urlprefix  [0]{URL }%
\providecommand \Eprint [0]{\href }%
\providecommand \doibase [0]{http://dx.doi.org/}%
\providecommand \selectlanguage [0]{\@gobble}%
\providecommand \bibinfo  [0]{\@secondoftwo}%
\providecommand \bibfield  [0]{\@secondoftwo}%
\providecommand \translation [1]{[#1]}%
\providecommand \BibitemOpen [0]{}%
\providecommand \bibitemStop [0]{}%
\providecommand \bibitemNoStop [0]{.\EOS\space}%
\providecommand \EOS [0]{\spacefactor3000\relax}%
\providecommand \BibitemShut  [1]{\csname bibitem#1\endcsname}%
\let\auto@bib@innerbib\@empty
\bibitem [{\citenamefont {Klinkhamer}(2019)}]{KI}%
  \BibitemOpen
  \bibfield  {author} {\bibinfo {author} {\bibfnamefont {F.~R.}\ \bibnamefont
  {Klinkhamer}},\ }\bibfield  {title} {\enquote {\bibinfo {title} {Regularized
  big bang singularity},}\ }\href {\doibase 10.1103/PhysRevD.100.023536}
  {\bibfield  {journal} {\bibinfo  {journal} {Phys.\ Rev.\ D}\ }\textbf
  {\bibinfo {volume} {100}},\ \bibinfo {pages} {023536} (\bibinfo {year}
  {2019})},\ \Eprint {http://arxiv.org/abs/1903.10450} {arXiv:1903.10450}
  \BibitemShut {NoStop}%
\bibitem [{\citenamefont {Klinkhamer}(2020{\natexlab{a}})}]{KII}%
  \BibitemOpen
  \bibfield  {author} {\bibinfo {author} {\bibfnamefont {F.~R.}\ \bibnamefont
  {Klinkhamer}},\ }\bibfield  {title} {\enquote {\bibinfo {title} {{More on the
  regularized big bang singularity}},}\ }\href {\doibase
  10.1103/PhysRevD.101.064029} {\bibfield  {journal} {\bibinfo  {journal}
  {Phys.\ Rev.\ D}\ }\textbf {\bibinfo {volume} {101}},\ \bibinfo {pages}
  {064029} (\bibinfo {year} {2020}{\natexlab{a}})},\ \Eprint
  {http://arxiv.org/abs/1907.06547} {arXiv:1907.06547} \BibitemShut {NoStop}%
\bibitem [{\citenamefont {Klinkhamer}\ and\ \citenamefont {Wang}(2019)}]{KWI}%
  \BibitemOpen
  \bibfield  {author} {\bibinfo {author} {\bibfnamefont {F.~R.}\ \bibnamefont
  {Klinkhamer}}\ and\ \bibinfo {author} {\bibfnamefont {Z.~L.}\ \bibnamefont
  {Wang}},\ }\bibfield  {title} {\enquote {\bibinfo {title} {{Nonsingular
  bouncing cosmology from general relativity}},}\ }\href {\doibase
  10.1103/PhysRevD.100.083534} {\bibfield  {journal} {\bibinfo  {journal}
  {Phys.\ Rev.\ D}\ }\textbf {\bibinfo {volume} {100}},\ \bibinfo {pages}
  {083534} (\bibinfo {year} {2019})},\ \Eprint
  {http://arxiv.org/abs/1904.09961} {arXiv:1904.09961} \BibitemShut {NoStop}%
\bibitem [{\citenamefont {Klinkhamer}\ and\ \citenamefont
  {Wang}(2020)}]{KWIII}%
  \BibitemOpen
  \bibfield  {author} {\bibinfo {author} {\bibfnamefont {F.~R.}\ \bibnamefont
  {Klinkhamer}}\ and\ \bibinfo {author} {\bibfnamefont {Z.~L.}\ \bibnamefont
  {Wang}},\ }\bibfield  {title} {\enquote {\bibinfo {title} {{Nonsingular
  bouncing cosmology from general relativity: Scalar metric perturbations}},}\
  }\href {\doibase 10.1103/PhysRevD.101.064061} {\bibfield  {journal} {\bibinfo
   {journal} {Phys.\ Rev.\ D}\ }\textbf {\bibinfo {volume} {101}},\ \bibinfo
  {pages} {064061} (\bibinfo {year} {2020})},\ \Eprint
  {http://arxiv.org/abs/1911.06173} {arXiv:1911.06173} \BibitemShut {NoStop}%
\bibitem [{\citenamefont
  {Klinkhamer}(2020{\natexlab{b}})}]{Klinkhamer:2020sxv}%
  \BibitemOpen
  \bibfield  {author} {\bibinfo {author} {\bibfnamefont {F.~R.}\ \bibnamefont
  {Klinkhamer}},\ }\href@noop {} {\enquote {\bibinfo {title} {Another model for
  the regularized big bang},}\ } (\bibinfo {year} {2020}{\natexlab{b}}),\
  \Eprint {http://arxiv.org/abs/2005.12157} {arXiv:2005.12157} \BibitemShut
  {NoStop}%
\bibitem [{\citenamefont {Ashtekar}(2009)}]{Ashtekar2008}%
  \BibitemOpen
  \bibfield  {author} {\bibinfo {author} {\bibfnamefont {A.}~\bibnamefont
  {Ashtekar}},\ }\bibfield  {title} {\enquote {\bibinfo {title} {{Loop Quantum
  Cosmology: An Overview}},}\ }\href {\doibase 10.1007/s10714-009-0763-4}
  {\bibfield  {journal} {\bibinfo  {journal} {Gen. Rel. Grav.}\ }\textbf
  {\bibinfo {volume} {41}},\ \bibinfo {pages} {707--741} (\bibinfo {year}
  {2009})},\ \Eprint {http://arxiv.org/abs/0812.0177} {arXiv:0812.0177}
  \BibitemShut {NoStop}%
\bibitem [{\citenamefont {Lidsey}\ \emph {et~al.}(2000)\citenamefont {Lidsey},
  \citenamefont {Wands},\ and\ \citenamefont {Copeland}}]{Lidsey1999}%
  \BibitemOpen
  \bibfield  {author} {\bibinfo {author} {\bibfnamefont {J.~E.}\ \bibnamefont
  {Lidsey}}, \bibinfo {author} {\bibfnamefont {D.}~\bibnamefont {Wands}}, \
  and\ \bibinfo {author} {\bibfnamefont {E.~J.}\ \bibnamefont {Copeland}},\
  }\bibfield  {title} {\enquote {\bibinfo {title} {{Superstring cosmology}},}\
  }\href {\doibase 10.1016/S0370-1573(00)00064-8} {\bibfield  {journal}
  {\bibinfo  {journal} {Phys. Rept.}\ }\textbf {\bibinfo {volume} {337}},\
  \bibinfo {pages} {343--492} (\bibinfo {year} {2000})},\ \Eprint
  {http://arxiv.org/abs/9909061} {arXiv:9909061} \BibitemShut {NoStop}%
\bibitem [{\citenamefont {Klinkhamer}(2020{\natexlab{c}})}]{K2020-IIB-3}%
  \BibitemOpen
  \bibfield  {author} {\bibinfo {author} {\bibfnamefont {F.~R.}\ \bibnamefont
  {Klinkhamer}},\ }\bibfield  {title} {\enquote {\bibinfo {title} {{IIB matrix
  model and regularized big bang}},}\ }\href {\doibase 10.1093/ptep/ptab059}
  {\bibfield  {journal} {\bibinfo  {journal} {PTEP}\ }\textbf {\bibinfo
  {volume} {2021}},\ \bibinfo {pages} {063} (\bibinfo {year}
  {2020}{\natexlab{c}})},\ \Eprint {http://arxiv.org/abs/2009.06525}
  {arXiv:2009.06525} \BibitemShut {NoStop}%
\bibitem [{\citenamefont {Klinkhamer}(2021)}]{Klinkhamer:2020wct}%
  \BibitemOpen
  \bibfield  {author} {\bibinfo {author} {\bibfnamefont {F.~R.}\ \bibnamefont
  {Klinkhamer}},\ }\bibfield  {title} {\enquote {\bibinfo {title} {{IIB matrix
  model: Emergent spacetime from the master field}},}\ }\href {\doibase
  10.1093/ptep/ptaa168} {\bibfield  {journal} {\bibinfo  {journal} {PTEP}\
  }\textbf {\bibinfo {volume} {2021}},\ \bibinfo {pages} {013B04} (\bibinfo
  {year} {2021})},\ \Eprint {http://arxiv.org/abs/2007.08485}
  {arXiv:2007.08485} \BibitemShut {NoStop}%
\bibitem [{\citenamefont {Klinkhamer}(2021, to appear in Acta Phys. Polon.
  B)}]{Klinkhamer2021a}%
  \BibitemOpen
  \bibfield  {author} {\bibinfo {author} {\bibfnamefont {F.~R.}\ \bibnamefont
  {Klinkhamer}},\ }\bibfield  {title} {\enquote {\bibinfo {title} {{M-theory
  and the birth of the Universe}},}\ }in\ \href@noop {} {\emph {\bibinfo
  {booktitle} {{27th Cracow Epiphany Conference on Future of particle physics
  }}}}\ (\bibinfo {year} {2021, to appear in Acta Phys. Polon. B})\ \Eprint
  {http://arxiv.org/abs/2102.11202} {arXiv:2102.11202} \BibitemShut {NoStop}%
\bibitem [{\citenamefont {Becker}\ \emph {et~al.}(2006)\citenamefont {Becker},
  \citenamefont {Becker},\ and\ \citenamefont {Schwarz}}]{Becker2006}%
  \BibitemOpen
  \bibfield  {author} {\bibinfo {author} {\bibfnamefont {K.}~\bibnamefont
  {Becker}}, \bibinfo {author} {\bibfnamefont {M.}~\bibnamefont {Becker}}, \
  and\ \bibinfo {author} {\bibfnamefont {J.~H.}\ \bibnamefont {Schwarz}},\
  }\href@noop {} {\emph {\bibinfo {title} {{String theory and M-theory: A
  modern introduction}}}}\ (\bibinfo  {publisher} {Cambridge University
  Press},\ \bibinfo {year} {2006})\BibitemShut {NoStop}%
\bibitem [{\citenamefont {G\"{u}nther}(September 2017)}]{Gunther}%
  \BibitemOpen
  \bibfield  {author} {\bibinfo {author} {\bibfnamefont {M.}~\bibnamefont
  {G\"{u}nther}},\ }\href@noop {} {\emph {\bibinfo {title} {Skyrmion spacetime
  defect, degenerate metric, and negative gravitational mass}}}\ (\bibinfo
  {publisher} {Master Thesis, KIT},\ \bibinfo {address} {Karlsruhe},\ \bibinfo
  {year} {September 2017})\BibitemShut {NoStop}%
\bibitem [{\citenamefont {Wald}(1984)}]{Wald}%
  \BibitemOpen
  \bibfield  {author} {\bibinfo {author} {\bibfnamefont {R.~M.}\ \bibnamefont
  {Wald}},\ }\href {https://cds.cern.ch/record/106274} {\emph {\bibinfo {title}
  {{General relativity}}}}\ (\bibinfo  {publisher} {Chicago University Press},\
  \bibinfo {address} {Chicago},\ \bibinfo {year} {1984})\BibitemShut {NoStop}%
\bibitem [{\citenamefont {Misner}\ \emph {et~al.}(1973)\citenamefont {Misner},
  \citenamefont {Thorne},\ and\ \citenamefont {Wheeler}}]{MTW}%
  \BibitemOpen
  \bibfield  {author} {\bibinfo {author} {\bibfnamefont {C.~W.}\ \bibnamefont
  {Misner}}, \bibinfo {author} {\bibfnamefont {K.~S.}\ \bibnamefont {Thorne}},
  \ and\ \bibinfo {author} {\bibfnamefont {J.~A.}\ \bibnamefont {Wheeler}},\
  }\href@noop {} {\emph {\bibinfo {title} {{Gravitation}}}}\ (\bibinfo
  {publisher} {W. H. Freeman},\ \bibinfo {address} {San Francisco},\ \bibinfo
  {year} {1973})\BibitemShut {NoStop}%
\bibitem [{\citenamefont {Akrami}\ \emph {et~al.}(2020)\citenamefont {Akrami}
  \emph {et~al.}}]{Planck2018}%
  \BibitemOpen
  \bibfield  {author} {\bibinfo {author} {\bibfnamefont {Y.}~\bibnamefont
  {Akrami}} \emph {et~al.} (\bibinfo {collaboration} {Planck}),\ }\bibfield
  {title} {\enquote {\bibinfo {title} {{Planck 2018 results. X. Constraints on
  inflation}},}\ }\href {\doibase 10.1051/0004-6361/201833887} {\bibfield
  {journal} {\bibinfo  {journal} {Astron. Astrophys.}\ }\textbf {\bibinfo
  {volume} {641}},\ \bibinfo {pages} {A10} (\bibinfo {year} {2020})},\ \Eprint
  {http://arxiv.org/abs/1807.06211} {arXiv:1807.06211} \BibitemShut {NoStop}%
\bibitem [{\citenamefont {Nojiri}\ \emph {et~al.}(2019)\citenamefont {Nojiri},
  \citenamefont {Odintsov}, \citenamefont {Oikonomou},\ and\ \citenamefont
  {Paul}}]{Odintsov2019}%
  \BibitemOpen
  \bibfield  {author} {\bibinfo {author} {\bibfnamefont {Shin'ichi}\
  \bibnamefont {Nojiri}}, \bibinfo {author} {\bibfnamefont {S.~D.}\
  \bibnamefont {Odintsov}}, \bibinfo {author} {\bibfnamefont {V.~K.}\
  \bibnamefont {Oikonomou}}, \ and\ \bibinfo {author} {\bibfnamefont {Tanmoy}\
  \bibnamefont {Paul}},\ }\bibfield  {title} {\enquote {\bibinfo {title}
  {{Nonsingular bounce cosmology from Lagrange multiplier $F(R)$ gravity}},}\
  }\href {\doibase 10.1103/PhysRevD.100.084056} {\bibfield  {journal} {\bibinfo
   {journal} {Phys. Rev. D}\ }\textbf {\bibinfo {volume} {100}},\ \bibinfo
  {pages} {084056} (\bibinfo {year} {2019})},\ \Eprint
  {http://arxiv.org/abs/1910.03546} {arXiv:1910.03546} \BibitemShut {NoStop}%
\bibitem [{\citenamefont {Elizalde}\ \emph {et~al.}(2020)\citenamefont
  {Elizalde}, \citenamefont {Odintsov}, \citenamefont {Oikonomou},\ and\
  \citenamefont {Paul}}]{Odintsov2020a}%
  \BibitemOpen
  \bibfield  {author} {\bibinfo {author} {\bibfnamefont {E.}~\bibnamefont
  {Elizalde}}, \bibinfo {author} {\bibfnamefont {S.~D.}\ \bibnamefont
  {Odintsov}}, \bibinfo {author} {\bibfnamefont {V.~K.}\ \bibnamefont
  {Oikonomou}}, \ and\ \bibinfo {author} {\bibfnamefont {Tanmoy}\ \bibnamefont
  {Paul}},\ }\bibfield  {title} {\enquote {\bibinfo {title} {{Extended matter
  bounce scenario in ghost free $f(R,\mathcal{G})$ gravity compatible with
  GW170817}},}\ }\href {\doibase 10.1016/j.nuclphysb.2020.114984} {\bibfield
  {journal} {\bibinfo  {journal} {Nucl. Phys. B}\ }\textbf {\bibinfo {volume}
  {954}},\ \bibinfo {pages} {114984} (\bibinfo {year} {2020})},\ \Eprint
  {http://arxiv.org/abs/2003.04264} {arXiv:2003.04264} \BibitemShut {NoStop}%
\bibitem [{\citenamefont {Odintsov}\ \emph {et~al.}(2020)\citenamefont
  {Odintsov}, \citenamefont {Oikonomou},\ and\ \citenamefont
  {Paul}}]{Odintsov2020b}%
  \BibitemOpen
  \bibfield  {author} {\bibinfo {author} {\bibfnamefont {S.~D.}\ \bibnamefont
  {Odintsov}}, \bibinfo {author} {\bibfnamefont {V.~K.}\ \bibnamefont
  {Oikonomou}}, \ and\ \bibinfo {author} {\bibfnamefont {Tanmoy}\ \bibnamefont
  {Paul}},\ }\bibfield  {title} {\enquote {\bibinfo {title} {{From a Bounce to
  the Dark Energy Era with $F(R)$ Gravity}},}\ }\href {\doibase
  10.1088/1361-6382/abbc47} {\bibfield  {journal} {\bibinfo  {journal} {Class.
  Quant. Grav.}\ }\textbf {\bibinfo {volume} {37}},\ \bibinfo {pages} {235005}
  (\bibinfo {year} {2020})},\ \Eprint {http://arxiv.org/abs/2009.09947}
  {arXiv:2009.09947} \BibitemShut {NoStop}%
\bibitem [{\citenamefont {Belinsky}\ \emph {et~al.}(1970)\citenamefont
  {Belinsky}, \citenamefont {Khalatnikov},\ and\ \citenamefont
  {Lifshitz}}]{BKL1970}%
  \BibitemOpen
  \bibfield  {author} {\bibinfo {author} {\bibfnamefont {V.~A.}\ \bibnamefont
  {Belinsky}}, \bibinfo {author} {\bibfnamefont {I.~M.}\ \bibnamefont
  {Khalatnikov}}, \ and\ \bibinfo {author} {\bibfnamefont {E.~M.}\ \bibnamefont
  {Lifshitz}},\ }\bibfield  {title} {\enquote {\bibinfo {title} {{Oscillatory
  approach to a singular point in the relativistic cosmology}},}\ }\href
  {\doibase 10.1080/00018737000101171} {\bibfield  {journal} {\bibinfo
  {journal} {Adv. Phys.}\ }\textbf {\bibinfo {volume} {19}},\ \bibinfo {pages}
  {525--573} (\bibinfo {year} {1970})}\BibitemShut {NoStop}%
\bibitem [{\citenamefont {Bunch}\ and\ \citenamefont
  {Davies}(1978)}]{Bunch1978}%
  \BibitemOpen
  \bibfield  {author} {\bibinfo {author} {\bibfnamefont {T.~S.}\ \bibnamefont
  {Bunch}}\ and\ \bibinfo {author} {\bibfnamefont {P.~C.~W.}\ \bibnamefont
  {Davies}},\ }\bibfield  {title} {\enquote {\bibinfo {title} {{Quantum Field
  Theory in de Sitter Space: Renormalization by Point Splitting}},}\ }\href
  {\doibase 10.1098/rspa.1978.0060} {\bibfield  {journal} {\bibinfo  {journal}
  {Proc. Roy. Soc. Lond. A}\ }\textbf {\bibinfo {volume} {360}},\ \bibinfo
  {pages} {117--134} (\bibinfo {year} {1978})}\BibitemShut {NoStop}%
\bibitem [{\citenamefont {Birrell}\ and\ \citenamefont
  {Davies}(1984)}]{Birrell1982}%
  \BibitemOpen
  \bibfield  {author} {\bibinfo {author} {\bibfnamefont {N.~D.}\ \bibnamefont
  {Birrell}}\ and\ \bibinfo {author} {\bibfnamefont {P.~C.~W.}\ \bibnamefont
  {Davies}},\ }\href {\doibase 10.1017/CBO9780511622632} {\emph {\bibinfo
  {title} {{Quantum Fields in Curved Space}}}},\ Cambridge Monographs on
  Mathematical Physics\ (\bibinfo  {publisher} {Cambridge Univ. Press},\
  \bibinfo {address} {Cambridge},\ \bibinfo {year} {1984})\BibitemShut
  {NoStop}%
\bibitem [{\citenamefont {Hartle}(2003)}]{Hartle2003gravity}%
  \BibitemOpen
  \bibfield  {author} {\bibinfo {author} {\bibfnamefont {J.~B.}\ \bibnamefont
  {Hartle}},\ }\href {https://books.google.de/books?id=ZHgpAQAAMAAJ} {\emph
  {\bibinfo {title} {Gravity: An Introduction to Einstein's General
  Relativity}}}\ (\bibinfo  {publisher} {Addison-Wesley},\ \bibinfo {address}
  {San Francisco},\ \bibinfo {year} {2003})\BibitemShut {NoStop}%
\bibitem [{\citenamefont {Manasse}\ and\ \citenamefont
  {Misner}(1963)}]{Manasse-Minser1963}%
  \BibitemOpen
  \bibfield  {author} {\bibinfo {author} {\bibfnamefont {F.~K.}\ \bibnamefont
  {Manasse}}\ and\ \bibinfo {author} {\bibfnamefont {C.~W.}\ \bibnamefont
  {Misner}},\ }\bibfield  {title} {\enquote {\bibinfo {title} {Fermi normal
  coordinates and some basic concepts in differential geometry},}\ }\href
  {\doibase 10.1063/1.1724316} {\bibfield  {journal} {\bibinfo  {journal}
  {Journal of Mathematical Physics}\ }\textbf {\bibinfo {volume} {4}},\
  \bibinfo {pages} {735--745} (\bibinfo {year} {1963})}\BibitemShut {NoStop}%
\bibitem [{\citenamefont {Nakahara}(2003)}]{Nakahara}%
  \BibitemOpen
  \bibfield  {author} {\bibinfo {author} {\bibfnamefont {M.}~\bibnamefont
  {Nakahara}},\ }\href {https://cds.cern.ch/record/206619} {\emph {\bibinfo
  {title} {{Geometry, topology and physics}}}},\ Graduate student series in
  physics\ (\bibinfo  {publisher} {Hilger},\ \bibinfo {address} {Bristol},\
  \bibinfo {year} {2003})\BibitemShut {NoStop}%
\bibitem [{\citenamefont {Gourgoulhon}(2012)}]{Gourgoulhon}%
  \BibitemOpen
  \bibfield  {author} {\bibinfo {author} {\bibfnamefont {E.}~\bibnamefont
  {Gourgoulhon}},\ }\href {https://books.google.de/books?id=XwB94Je8nnIC}
  {\emph {\bibinfo {title} {3+1 Formalism in General Relativity: Bases of
  Numerical Relativity}}},\ Lecture Notes in Physics\ (\bibinfo  {publisher}
  {Springer},\ \bibinfo {address} {Berlin},\ \bibinfo {year}
  {2012})\BibitemShut {NoStop}%
\bibitem [{\citenamefont {Choquet-Bruhat}(2014)}]{Choquet-Bruhat2014}%
  \BibitemOpen
  \bibfield  {author} {\bibinfo {author} {\bibfnamefont {Y.}~\bibnamefont
  {Choquet-Bruhat}},\ }\href
  {https://books.google.de/books/about/Introduction_to_General_Relativity_Black.html?id=rOYwBQAAQBAJ&redir_esc=y}
  {\emph {\bibinfo {title} {{Introduction to General Relativity, Black Holes,
  and Cosmology}}}}\ (\bibinfo  {publisher} {Oxford University Press},\
  \bibinfo {address} {Oxford},\ \bibinfo {year} {2014})\BibitemShut {NoStop}%
\bibitem [{\citenamefont {Hawking}\ and\ \citenamefont
  {Ellis}(1973)}]{Hawking-Ellis}%
  \BibitemOpen
  \bibfield  {author} {\bibinfo {author} {\bibfnamefont {S.~W.}\ \bibnamefont
  {Hawking}}\ and\ \bibinfo {author} {\bibfnamefont {G.~F.~R.}\ \bibnamefont
  {Ellis}},\ }\href {\doibase 10.1017/CBO9780511524646} {\emph {\bibinfo
  {title} {{The Large Scale Structure of Space-Time}}}},\ Cambridge Monographs
  on Mathematical Physics\ (\bibinfo  {publisher} {Cambridge University
  Press},\ \bibinfo {address} {Cambridge},\ \bibinfo {year} {1973})\BibitemShut
  {NoStop}%
\bibitem [{\citenamefont {Bondi}(1957)}]{Bondi1957}%
  \BibitemOpen
  \bibfield  {author} {\bibinfo {author} {\bibfnamefont {H.}~\bibnamefont
  {Bondi}},\ }\bibfield  {title} {\enquote {\bibinfo {title} {{Negative Mass in
  General Relativity}},}\ }\href {\doibase 10.1103/RevModPhys.29.423}
  {\bibfield  {journal} {\bibinfo  {journal} {Rev. Mod. Phys.}\ }\textbf
  {\bibinfo {volume} {29}},\ \bibinfo {pages} {423--428} (\bibinfo {year}
  {1957})}\BibitemShut {NoStop}%
\bibitem [{\citenamefont {Klinkhamer}\ and\ \citenamefont
  {Queiruga}(2018)}]{Klinkhamer-Queiruga2018}%
  \BibitemOpen
  \bibfield  {author} {\bibinfo {author} {\bibfnamefont {F.~R.}\ \bibnamefont
  {Klinkhamer}}\ and\ \bibinfo {author} {\bibfnamefont {J.~M.}\ \bibnamefont
  {Queiruga}},\ }\bibfield  {title} {\enquote {\bibinfo {title} {{Antigravity
  from a spacetime defect}},}\ }\href {\doibase 10.1103/PhysRevD.97.124047}
  {\bibfield  {journal} {\bibinfo  {journal} {Phys. Rev. D}\ }\textbf {\bibinfo
  {volume} {97}},\ \bibinfo {pages} {124047} (\bibinfo {year} {2018})},\
  \Eprint {http://arxiv.org/abs/1803.09736} {arXiv:1803.09736} \BibitemShut
  {NoStop}%
\bibitem [{\citenamefont {Farnes}(2018)}]{Farnes2017}%
  \BibitemOpen
  \bibfield  {author} {\bibinfo {author} {\bibfnamefont {J.~S.}\ \bibnamefont
  {Farnes}},\ }\bibfield  {title} {\enquote {\bibinfo {title} {{A unifying
  theory of dark energy and dark matter: Negative masses and matter creation
  within a modified $\Lambda$CDM framework}},}\ }\href {\doibase
  10.1051/0004-6361/201832898} {\bibfield  {journal} {\bibinfo  {journal}
  {Astron. Astrophys.}\ }\textbf {\bibinfo {volume} {620}},\ \bibinfo {pages}
  {A92} (\bibinfo {year} {2018})},\ \Eprint {http://arxiv.org/abs/1712.07962}
  {arXiv:1712.07962} \BibitemShut {NoStop}%
\end{thebibliography}%


%

\end{document}